\documentclass[notitlepage,letterpaper,aps,prd,twocolumn,superscriptaddress,
nofootinbib,longbibliography,showpacs]{revtex4-1}
\pdfoutput=1
\usepackage{amsmath}
\usepackage{amsfonts}
\usepackage{amssymb}
\usepackage[latin1,utf8]{inputenc}
\usepackage{xcolor,cancel}
\usepackage[normalem]{ulem}
\RequirePackage{flushend}

\usepackage[colorlinks=true, pdfstartview = FitH, bookmarksopen = true, bookmarksopenlevel=2]{hyperref}
\usepackage{hyperref}
\hypersetup{
	colorlinks=true,
	linkcolor=blue,
	filecolor=magenta,      
	urlcolor=blue,
	pdftitle={DS_fQ},
	pdfpagemode=FullScreen,
}

\usepackage[all]{hypcap}

\usepackage{latexsym}
\usepackage{graphicx}
\usepackage{color}
\usepackage{subfigure}

\begin{document}
	
	\title{Cosmology in $f(Q)$ gravity: A unified dynamical 
system analysis at background and perturbation levels}

	\author{Wompherdeiki Khyllep}
	\affiliation{Department of Mathematics,
		St.\ Anthony's College, Shillong, Meghalaya 793001, India}
		
	\author{Jibitesh Dutta}
	\affiliation{Mathematics Division, Department of Basic Sciences and Social
		Sciences, North-Eastern Hill University,  Shillong, Meghalaya 793022, India}
	\affiliation{Inter University Centre for Astronomy and Astrophysics, Pune
		411007, India }

		\author{Emmanuel N. Saridakis}
\affiliation{National Observatory of Athens, Lofos Nymfon, 11852 Athens,
Greece}
\affiliation{CAS Key Laboratory for Researches in Galaxies and Cosmology,
Department of Astronomy, University of Science and Technology of China, Hefei,
Anhui 230026, P.R. China}
\affiliation{School of Astronomy, School of Physical Sciences,
University of Science and Technology of China, Hefei 230026, P.R. China}

\author{Kuralay Yesmakhanova}
\affiliation{Ratbay Myrzakulov Eurasian International Centre for Theoretical 
Physics, Nur-Sultan 010009,  Kazakhstan}
 \affiliation{Eurasian National University, Nur-Sultan Astana 010008, 
Kazakhstan}

	\begin{abstract}
		 
 Motivated by the fact that cosmological models based on $f(Q)$ 
 gravity are very efficient in fitting observational datasets at both 
 background and perturbation levels,  we perform a  combined dynamical system analysis of both background and perturbation equations in order to examine the validity of this result through an independent method. We examine two studied $f(Q)$ models of the literature, namely the power-law and the exponential ones. For both cases, we obtain a matter-dominated saddle point characterized by the correct growth rate of matter perturbations, followed by the transition to a stable dark-energy dominated accelerated universe in which matter perturbations remain constant. Furthermore,   analyzing the behavior of $f \sigma_8$, we find that the models fit observational data successfully, obtaining a behavior similar to that of $\Lambda$CDM scenario, although the exponential model does not possess the latter as a  particular limit. Hence, through the independent approach of dynamical systems, we do verify  the results of observational confrontation, namely that $f(Q)$ gravity can be considered as a very promising alternative to the $\Lambda$CDM concordance model.  
		
	\end{abstract}
	
	\maketitle

 \section{Introduction}\label{sec:intro}
	
Einstein's theory of General Relativity (GR) is the most successful theory 
to describe the  gravitational interaction, and based on that, the  Lambda Cold Dark Matter ($\Lambda$CDM) scenario  is the concordance    
cosmological model. However, this standard gravitational and cosmological paradigm faces  theoretical and observational problems 
such as the non-renormalizability of GR, the  cosmological 
constant problem, the coincidence problem, the Hubble tension, the $\sigma_8$ tension, etc \cite{Addazi:2021xuf,Abdalla:2022yfr,Martin:2012bt,
	Freedman:2017yms, Lusso:2019akb,Lin:2019htv,Perivolaropoulos:2021jda}. 	Hence, 
in the literature, one may find various modified theories of gravity 
\cite{CANTATA:2021ktz,Capozziello:2011et,Clifton:2011jh,Cai:2015emx} and several GR-based models beyond the $\Lambda$CDM one  
\cite{Copeland:2006wr,Joyce:2016vqv,Cai:2009zp}, that aim to alleviate a part of or all the above issues.
	
The usual way to construct gravitational modifications is 
to add extra terms in the Einstein-Hilbert Lagrangian, 
resulting to    $f(R)$ gravity 
\cite{Starobinsky:1980te,Capozziello:2002rd,DeFelice:2010aj},   Gauss-Bonnet 
and $f(G)$
gravity \cite{Antoniadis:1993jc,Nojiri:2005jg,DeFelice:2008wz},   cubic and 
$f(P)$ 
gravity 
\cite{Erices:2019mkd,Marciu:2020ysf,BeltranJimenez:2020lee},    
Horndeski/Galileon scalar-tensor theories 
\cite{Horndeski:1974wa,Deffayet:2009wt},  etc. Alternatively, one can add new 
terms to the  equivalent   
formulation of gravity based on torsion, resulting to  $f(T)$ gravity 
\cite{Bengochea:2008gz,Cai:2015emx},   $f(T,T_{G})$ gravity 
\cite{Kofinas:2014owa,Kofinas:2014aka, Kofinas:2014daa},   $f(T,B)$ gravity 
\cite{Bahamonde:2015zma,Bahamonde:2016grb},
 scalar-torsion
gravity \cite{Geng:2011aj}, etc.   Nevertheless, there is a third way to 
construct new classes of modified gravity,  starting  from the  
``symmetric 
teleparallel gravity'', which is based on the non-metricity scalar $Q$  
\cite{BeltranJimenez:2017tkd}, and extend it 
to  a function  $f(Q)$ in the Lagrangian. 

The modified theory of $f(Q)$ gravity leads to interesting cosmological
phenomenology at the background level \cite{Jimenez:2019ovq, Dialektopoulos:2019mtr, Bajardi:2020fxh, Flathmann:2020zyj,Mandal:2020buf,DAmbrosio:2020nev,Mandal:2020lyq, Dimakis:2021gby, Nakayama:2021rda, Khyllep:2021pcu,Hohmann:2021ast,Wang:2021zaz,Quiros:2021eju, 
	Ferreira:2022jcd,Solanki:2022ccf,De:2022shr, Solanki:2021qni, 
	Capozziello:2022wgl,Narawade:2022jeg,Dimakis:2022rkd,Albuquerque:2022eac,Arora:2022mlo,Pati:2022dwl}. Additionally, it  has been successfully confronted with various background and perturbation observational data, such as the Cosmic Microwave Background (CMB), 
Supernovae type Ia (SNIa), Baryonic Acoustic Oscillations (BAO), Redshift Space Distortion (RSD), growth data, etc, \cite{Soudi:2018dhv,Lazkoz:2019sjl,Barros:2020bgg,Ayuso:2020dcu, 
	Anagnostopoulos:2021ydo,Mandal:2021bpd,Atayde:2021pgb,Frusciante:2021sio}, and this confrontation reveals that $f(Q)$ gravity 
may challenge the standard $\Lambda$CDM scenario. Finally, 
$f(Q)$ gravity  comfortably passes the Big Bang Nucleosynthesis (BBN) constraints too \cite{Anagnostopoulos:2022gej}.

Motivated by the exciting features of  $f(Q)$ gravity, in this work, we employ the powerful mathematical tool of dynamical system analysis in order to investigate for the first time the cosmological dynamics of $f(Q)$ cosmology at both background and perturbation levels. Such investigation can be used to further confirm the results obtained from the observational analysis. We mention that usually, the dynamical system approach is   applied  at the background level  \cite{wainwrightellis1997,Coley:2003mj,Bahamonde:2017ize,Copeland:1997et,	Gong:2006sp, Setare:2008sf, Matos:2009hf, Copeland:2009be,Leyva:2009zz,Leon:2010pu, Urena-Lopez:2011gxx, 
	Leon:2013qh,Fadragas:2013ina, Skugoreva:2014ena,Dutta:2016bbs, 
	Dutta:2017kch, Zonunmawia:2018xvf,Khyllep:2021yyp}, however relatively 
recently it was realized that the analysis can be applied at the perturbation level  too \cite{Basilakos:2019dof,Alho:2019jho,Landim:2019lvl,Khyllep:2021wjd}.  
Hence, with this combined analysis, we can determine both the background stable late-time solutions, as well as the growth of the structure formation, independent of the specific initial conditions. Moreover, we can examine 
how the matter perturbations back-react to  the  background 
solutions, too. 

The work is organized as follows: In Sec. \ref{sec:f_Q_comology}, we present the field equations of $f(Q)$ gravity, from which one can arrive at the background and perturbed cosmological equations. Sec. \ref{sec:DSA} deals with the dynamical analysis of the combined system for the power-law and exponential models. Finally, the obtained results are summarized in Sec. \ref{sec:conc}.

	\section{$f(Q)$ cosmology}\label{sec:f_Q_comology}
	
	In this section we briefly review $f(Q)$ cosmology. 
	The action of   $f(Q)$ gravity is
	given by \cite{BeltranJimenez:2017tkd,Jimenez:2019ovq}
	\begin{equation}  \label{action00}
		S=\int \left[ -\frac{1}{16\pi G}f(Q)+\mathcal{L}_m\right]\sqrt{-g}~d^4x,
	\end{equation}
	where  $g$ is the
	determinant of the metric $g_{\mu\nu}$ and $\mathcal{L}_m$ is the matter
	Lagrangian density. 	
	$f(Q)$ is an arbitrary function of the  non-metricity scalar  
\cite{BeltranJimenez:2017tkd}  
  \begin{equation}
\label{NontyScalar}
Q=-\frac{1}{4}Q_{\alpha \beta \gamma}Q^{\alpha \beta 
\gamma}+\frac{1}{2}Q_{\alpha \beta \gamma}Q^{ \gamma \beta 
\alpha}+\frac{1}{4}Q_{\alpha}Q^{\alpha}-\frac{1}{2}Q_{\alpha}\tilde{Q}^{\alpha} 
\,,
\end{equation}
where
$Q_{\alpha}\equiv Q_{\alpha \ \mu}^{\ \: \mu} \,   $ and
$\tilde{Q}^{\alpha} 
\equiv Q_{\mu }^{\ \: \mu \alpha}  \,  
$ are acquired  from contractions of the  non-metricity tensor $
    Q_{\alpha\mu\nu}\equiv\nabla_\alpha g_{\mu\nu}$. Thus, 
 Symmetric Teleparallel Equivalent of General 
Relativity, and therefore General Relativity, is 
recovered for $f(Q)=Q$.

Variation of  action (\ref{action00}), and setting $8\pi G=1$ for simplicity,
leads to the field equations 
\cite{Jimenez:2019ovq, Dialektopoulos:2019mtr}: 
\begin{eqnarray}
&&  
\frac{2}{\sqrt{-g}} \nabla_{\alpha}\left\{\sqrt{-g} g_{\beta \nu} f_{Q} 
\left[- \frac{1}{2} L^{\alpha \mu \beta}+ \frac{1}{4} g^{\mu \beta} 
\left(Q^\alpha -  \tilde{Q}^\alpha \right) \right.\right.\nonumber\\
&&\left.\left. \ \ \ \ \ \ \ \ \ \ \ \ \ \ \ \ \ \  \ \ \ \ \ \ \ \ \ \ \ \, 
- \frac{1}{8} 
\left(g^{\alpha \mu} Q^\beta + g^{\alpha \beta} Q^\mu  
\right)\right]\right\}
 \nonumber \\
&& + f_{Q} \left[- \frac{1}{2} L^{\mu \alpha \beta}- \frac{1}{8} \left(g^{\mu 
\alpha} Q^\beta 
+ g^{\mu \beta} Q^\alpha  \right)
 \right. \nonumber\\
&&\left. \ \ \ \ \ \ \ + \frac{1}{4} g^{\alpha 
\beta} \left(Q^\mu -  \tilde{Q}^\mu \right)
\right] Q_{\nu \alpha 
\beta} +\frac{1}{2} \delta_{\nu}^{\mu} f=T_{\,\,\,\nu}^{\mu}\,,
\label{eoms}
\end{eqnarray}
where 
$
L^{\alpha}_{\,\,\mu\nu}=\frac{1}{2}Q^{\alpha}_{\,\,\mu\nu}-Q^{\,\,\,\alpha}_{
(\mu\,\,\,\nu)} $ is  the  disformation tensor, $T_{\mu\nu}$ is the matter 
energy-momentum 
tensor, and
$f_{Q}\equiv\partial f/\partial Q$. 

 At the background level, we assume a  homogeneous,  isotropic and spatially 
flat Friedmann-Lema\^itre-Robertson-Walker (FLRW) spacetime, whose metric is of 
the form
	\begin{equation}\label{eq:FLRW}
		ds^{2}=-dt^{2}+a^{2}(t)(dx^2+dy^2+dz^2)\,,
	\end{equation}%
	where $t$ is the cosmic time, $a(t)$ is the scale factor and $x, y, z$ are 
the Cartesian coordinates. Note that in FLRW metric, for the non-metricity 
scalar we obtain $Q=6H^{2}$, where  
$H=\frac{\dot{a}}{a}$ is the Hubble function and the upper dot denotes 
derivative with respect to   $t$.  
Imposing the splitting   $f(Q)=Q+F(Q)$, and 
applying  the FLRW metric, the  corresponding Friedman equations are 
\cite{BeltranJimenez:2017tkd,Jimenez:2019ovq}
	\begin{eqnarray}
		3H^{2}=\rho+\frac{F}{2}-QF_Q \,,\label{eq:FRDEQ1}\\
		\left(2Q F_{QQ}+F_Q+1\right) \dot{H}+\frac{1}{4} \left(Q+2QF_Q-F\right)&=&-2p\,,~~~\label{eq:FRDEQ2}
	\end{eqnarray}
	with    $F_Q\equiv\frac{dF}{dQ}$,  $F_{QQ}\equiv\frac{d^2F}{dQ^2}$. In the 
above equations   
$\rho$ and $p$ are the energy density and pressure of the matter fluid, which 
for no  interaction satisfy the conservation equation   
	\begin{align}
		\dot{\rho}+3H(1+w)\rho =0\,,\label{eq:cons}
	\end{align}
	with  $w\equiv p/\rho$   the matter equation-of-state parameter.

We can now introduce the effective, total, energy density $\rho 
_{\mathrm{eff}}$ and 	  pressure $p_{\mathrm{eff}}$, respectively as
	\begin{eqnarray}
		\rho _{\mathrm{eff}} &\equiv&\rho+\frac{F}{2}-QF_{Q}\,, \\
		p_{\mathrm{eff}} &\equiv&\frac{\rho (1+w)}{2QF_{QQ}+F_Q+1}-\frac{Q}{2} 
\,,
	\end{eqnarray}%
	and thus the corresponding   total  equation of state 
$w_{\mathrm{eff}}$ is given by
	\begin{equation}
		w_{\mathrm{eff}}\equiv\frac{p_{\mathrm{eff}}}{\rho 
_{\mathrm{eff}}}=-1+\frac{\Omega _{m}(1+w)}{2QF_{QQ}+F_{Q}+1}\,.
	\end{equation}
	We mention that for an accelerated Universe one requires
$w_{\mathrm{eff}}<-\frac{1}{3}$. 
Finally, it proves convenient to 
  introduce the energy density parameters  
as
	\begin{equation}		
\Omega_{m}=\frac{\rho}{3H^{2}},~~~~~~\Omega_{Q}=\frac{\frac{F}{2}-QF_Q}{3H^{2}}\
,
	\end{equation}%
	 and thus the first Friedman equation  \eqref{eq:FRDEQ1} becomes simply
	\begin{equation}
		\Omega _{m}+\Omega _{Q}=1\,.  \label{constraint}
	\end{equation}%
	
Let us proceed to the investigation of the linear perturbation level, focusing 
on the   matter density contrast $\delta=\frac{\delta \rho}{\rho}$, where 
$\delta \rho$ is the perturbation of the matter energy density. In particular, 
the evolution equation of the matter overdensity  at the quasi-static limit  is 
given by \cite{Jimenez:2019ovq,Anagnostopoulos:2021ydo}	
	\begin{equation}\label{eq:overdens}
		\ddot{\delta}+ 2H \dot{\delta}=\frac{\rho \delta}{2(1+F_{Q})}\,,
	\end{equation}
	where the denominator of the last term accounts for the appearance of an 
effective Newton's constant.
		We mention that  in the quasi-static limit the terms involving time 
derivatives in the perturbed equations are neglected, and only spatial 
derivative terms remain. It is worth mentioning that such an approximation is 
sufficient at small scales, well within the cosmic horizon 
\cite{Song:2006ej}.

	\section{Dynamical system analysis}
	\label{sec:DSA}

	In this section  we construct the dynamical system of the background and 
perturbed equations, for a general function $F(Q)$.  In this regard, we 
transform the equations \eqref{eq:FRDEQ1}-\eqref{eq:cons} and 
\eqref{eq:overdens} into a first-order autonomous system, by considering the 
following dynamical variables:
	\begin{align}\label{eq:dyn_var}
		x=\frac{F}{6H^2},~~~~y=-2F_Q,~~~~ u=\frac{d(\ln \delta)}{d(\ln a)}\,.
	\end{align}
	Hence,  while the variables $x, y$ are associated with the background
	evolution of the Universe, the variable $u$ quantifies the growth of matter 
perturbations. Therefore, $u>0$ signifies the growth and
	$u<0$ indicates the decay of matter perturbations, whenever
	the matter density contrast $\delta$ is positive.
	
	The background cosmological parameters $\Omega_{m}$, $\Omega_{Q}$ and 
$w_{\rm eff}$ can  expressed as
	\begin{eqnarray}
		&&\Omega_{m}=1-x-y,\nonumber\\
		&&\Omega_{Q}=x+y,\nonumber\\ 
		&&w_{\rm 
eff}=-1+\frac{(1-x-y)(1+w)}{2QF_{QQ}-\frac{y}{2}+1}\,.
	\end{eqnarray}
	Now, in terms of variables \eqref{eq:dyn_var}, the cosmological 
equations can be written as the following  dynamical system:
	\begin{eqnarray}
		x' &=& -\frac{\dot{H}}{H^2} (y+2x)\,,\label{eq:xp}\\
		y'&=& 2 y \frac{\dot{H}}{H^2} \frac{Q F_{QQ}}{F_Q}\,,\label{eq:yp}\\
		u'&=& -u(u+2)+\frac{3(1-x-y)}{ (2-y)}-\frac{\dot{H}}{H^2} u\,,\label{eq:up}
	\end{eqnarray}
	where a prime stands for differentiation with respect to $\ln a$, and 
\begin{eqnarray}\frac{\dot{H}}{H^2}=-\frac{3-3(x+y)}{4QF_{QQ}-y+2}.
 \end{eqnarray}
 The 
physical system is a product space  of the background phase space $\mathbb{B}$, 
which includes the variables $x, y$, and the perturbed space $\mathbb{P}$ 
which 
consists of the variable $u$. Under the physical condition $0 \leq \Omega_{m} 
\leq 1$, the phase space of the combined system is 
	\begin{equation}
		\Psi=\mathbb{B} \times \mathbb{P}=\left\{(x,y,u) \in \mathbb{R}^2 \times \mathbb{R}: 0\leq x+y \leq 1\right\}\,.
	\end{equation}
		It is worth mentioning that the projection of orbits of the product 
space $\Psi$ on    space 	$\mathbb{B}$ reduces to the corresponding 
background 
orbits.
	
 As a next step we shall determine the dynamical evolution of the system by 
extracting its critical points  and examining their stability. Physically, a 
stable point with $u>0$ indicates an indefinite growth of matter perturbations 
and hence  the system is not stable with respect to matter perturbations. 
However, a stable point with $u < 0$ indicates the decay of matter perturbations 
and therefore the system is asymptotically stable with respect to
	  perturbations. Finally, when $u=0$ at a stable
	point  implies that   matter perturbations  remain constant. In summary, 
for a viable model  one desires to have an unstable or saddle point with $u>0$, 
required for the description of the matter epoch of the universe, in which the 
matter perturbations grow but which does not hold eternally, followed by a 
stable late-time attractor corresponding to acceleration but with $u=0$.

In order to proceed to specific analysis, we need to specify the function  $F$ 
and hence determine the term $\frac{QF_{QQ}}{F_Q}$. In the following 
subsections we will consider two specific models, which are known to lead to 
interesting cosmological phenomenology.

\subsection{Model I: $F(Q)=\alpha 
\left(\frac{Q}{Q_0}\right)^n$}\label{sec:pow_model}
	
 We start our analysis by considering a power-law model with 
\cite{Lazkoz:2019sjl,Ayuso:2020dcu,Jimenez:2019ovq}
 	\begin{eqnarray}
 	\label{ModelI}
F(Q)=\alpha 
\left(\frac{Q}{Q_0}\right)^n,
 	\end{eqnarray} 
with	 $\alpha$ and $n$  two   parameters and where $Q_0=6H_0^2$ is the 
present value of $Q$ (note that applying the first Friedmann equation at 
present, $\alpha$ can be eliminated in terms of $n$ and the present value 
$\Omega_{m0}$). This model can describe the late-time universe 
acceleration and it is also compatible with   BBN constraints 
\cite{Anagnostopoulos:2022gej}.
We mention that   for $n=0$  this model is 
equivalent to the concordance $\Lambda$CDM scenario, while for $n=1$ the model 
reduces to the symmetric teleparallel equivalent of general relativity 
\cite{Lazkoz:2019sjl,Ayuso:2020dcu,Jimenez:2019ovq}.  
 In this case, we have  $QF_{QQ}=\frac{(1-n)y}{4}$ and hence the system 
\eqref{eq:xp}-\eqref{eq:up} closes. 

The corresponding dynamical system contains the 
following four critical points:

	\begin{itemize}
		\item {\it Point $A_1$ $(0,0,1)$}:  This point corresponds to a 
matter-dominated critical solution with the background parameters $\Omega_{m}=1$ 
and $w_{\rm eff}=0$. At the perturbation level  we have $u=1$, which implies 
that the matter overdensity $\delta$  varies as the scale factor $a$ and hence 
increases with the universe expansion. The corresponding Jacobian matrix 
  has the eigenvalues $-\frac{5}{2}, 3$ and $\frac{3}{2} 
(1-n)$, therefore  point $A_1$  for any value of $n$  is always a saddle one. 
Hence, the trajectories pass through this point and leave it  as they are 
attracted by   a  late-time stable point. Thus, we   conclude that 
this point could be the main candidate for describing the structure formation 
during the matter domination era at both the background and perturbation 
levels. 
		
		\item {\it  Point $B_1$ $\left( 0,0,-\frac{3}{2}\right)$}: At the 
background level  this point corresponds to   matter domination, with 
$\Omega_{m}=1$ and $w_{\rm eff}=0$. However, this point could not  describe the 
formation of structures at the perturbation level, since $u=-\frac{3}{2}$, and 
hence the matter overdensity $\delta$  varies as $a^{-\frac{3}{2}}$. The 
eigenvalues of the  Jacobian matrix  are $\frac{5}{2}, 3$ and 
$\frac{3}{2} (1-n)$, and therefore this point is unstable for $n<1$ and saddle 
for $n>1$.  
		
		\item {\it 
		The curve of critical points $C_1$ $(1-y,y,0)$}: Each point on this 
curve corresponds to a solution dominated by the effective dark-energy 
  component, i.e. $\Omega_{Q}=1$, in which   the universe is 
  accelerated   with a cosmological-constant-like behavior, namely with  
$w_{\rm eff}=-1$. Furthermore, at the perturbation level  we have $u=0$, which 
implies that the matter perturbation remains constant. The corresponding
  eigenvalues are $-2, -3$ and $0$, and since the curve  is 
one-dimensional with one vanishing eigenvalue, it is   normally 
hyperbolic \cite{Coley:2003mj}, and one can determine its stability  by
examining  the signature of the remaining non-vanishing eigenvalues  
\cite{Coley:2003mj}. Therefore, we  conclude that it is
always stable. In summary, the curve $C_1$ describes the late 
time dark-energy   dominated Universe, at both   background and 
perturbation levels.
		
		\item {\it Curve of critical points $D_1$ $(1-y,y,-2)$}: Similarly to  
 $C_1$, this curve of critical points also corresponds to a 
cosmological-constant-like solution, i.e with  $w_{\rm eff}=-1$, dominated by 
the effective dark-energy component. Additionally, it is characterized by 
 the decay of matter perturbations, since $u=-2$. However, it is  saddle 
  with eigenvalues $2, -3$ 
and $0$. Hence, unlike  $C_1$, curve $D_1$ cannot describe a late-time 
 dark-energy dominated Universe at the perturbation level.
		
	\end{itemize}

	\begin{figure}[ht]
		\centering
		\includegraphics[width=7cm, height=7.2cm]{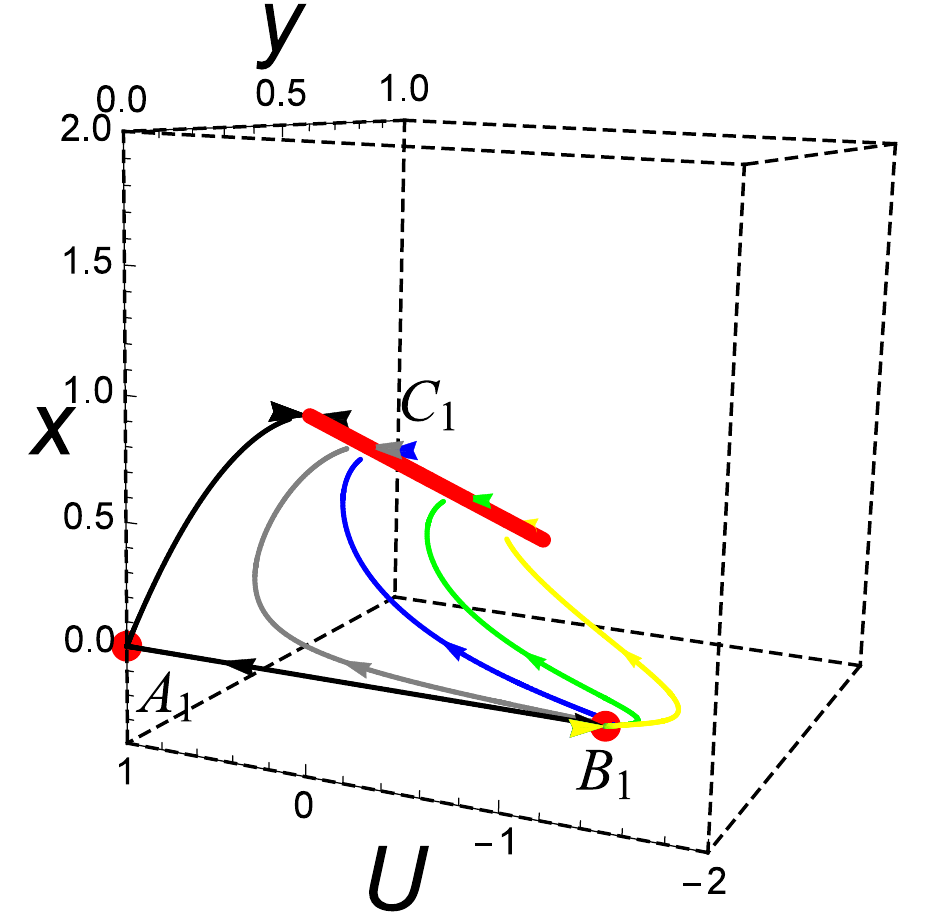} 
\label{fig:fig1_pow_phase}
		\caption{{\it{The phase portrait of the system 
\eqref{eq:xp}-\eqref{eq:up}, for the   power-law model I of (\ref{ModelI})
with $n=0.5$. This particular example exhibits  the evolution $B_1 \to 
A_1 \to C_1$.}}}
		\label{fig:fig1_pow}
	\end{figure}

	Our analysis reveals that different critical points describe different 
modes of matter perturbations. Additionally, we mention that identical 
background critical points behave differently at the perturbation level. For 
instance, we showed that points $A_1$ and $B_1$ describe the decelerated 
matter-dominated era at the background level, but only   point $A_1$ has the 
correct growth of matter structure. Interestingly, point $A_1$ is  saddle and 
thus it provides the natural exit towards a late-time accelerated epoch. On 
the other hand, at late times the curves of critical points  $C_1$ and $D_1$ 
are identical at the background level, describing the accelerated  dark-energy 
dominated epoch. However, only curve $C_1$  is physically and observationally 
interesting at the perturbation level, since it is   stable  and exhibits  
constant matter perturbations.  Lastly,     examining for completeness whether 
there are  critical points 
at infinity, we find that  no  such physical  points exist.

\begin{figure}[ht]
		\centering
			\includegraphics[width=6.cm, height=6.cm]{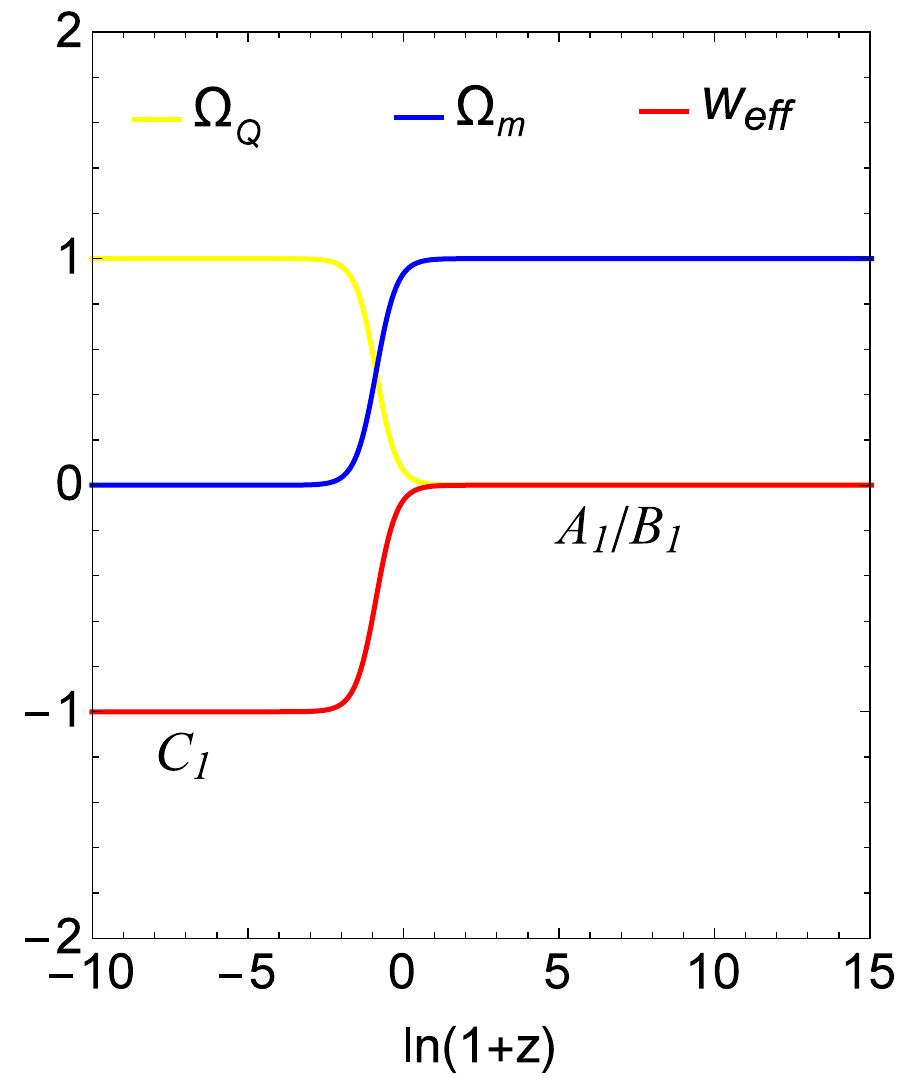}
			\includegraphics[width=6.cm, height=6.cm]{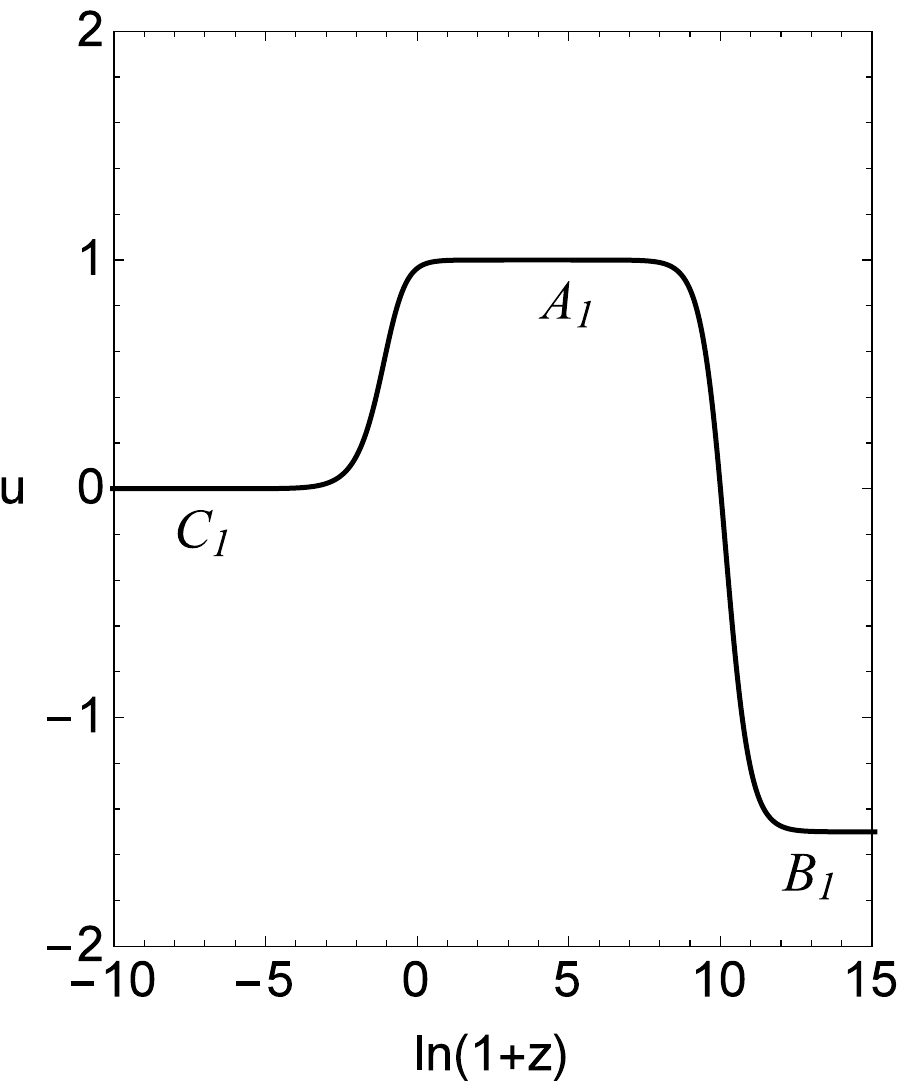} 
		\caption{
		{\it{ Upper graph:   Evolution of the density parameters of matter   	
$\Omega_{m}$  and of  non-metricity (dark-energy) $\Omega_{Q}$, 
 as well as of the total, effective 
equation-of-state parameter $w_{\rm eff}$, as functions of the redshift, for   
the   power-law model I of (\ref{ModelI}) with $n=0.5$.   Lower graph:   
Evolution     of the perturbation variable $u=\frac{d(\ln \delta)}{d(\ln a)}$. 
}}}
		\label{fig:fig2_pow}
	\end{figure}

In order to give the above information in a more transparent way, 	
  we display the phase portrait  of the system \eqref{eq:xp}-\eqref{eq:up} in 
Fig. \ref{fig:fig1_pow_phase}. As we see, the system   follows the orbit $B_1 
\to A_1 \to C_1$. Furthermore, in 
Fig. \ref{fig:fig2_pow}  we present the evolution
	of the background parameters and the matter growth rate  variable $u$,
	in terms of the  redshift 
$z=\frac{a_0}{a}-1$ with	$a_0 = 1$   the current scale factor.
As we see, the model describes the  transition from   matter  domination towards 
an accelerated dark-energy dominated epoch.

In summary, the present power-law model can describe the desired thermal 
history of the Universe, both at the background and perturbation levels. 
Our analysis indicates that in principle the above hold  for any value 
of $n$. Nevertheless, we should mention  that a  tuning of initial 
conditions is required in order to have a sufficiently long matter-dominated 
era.

\begin{figure}[ht]
		\centering
		\includegraphics[width=8cm, height=7.2cm]{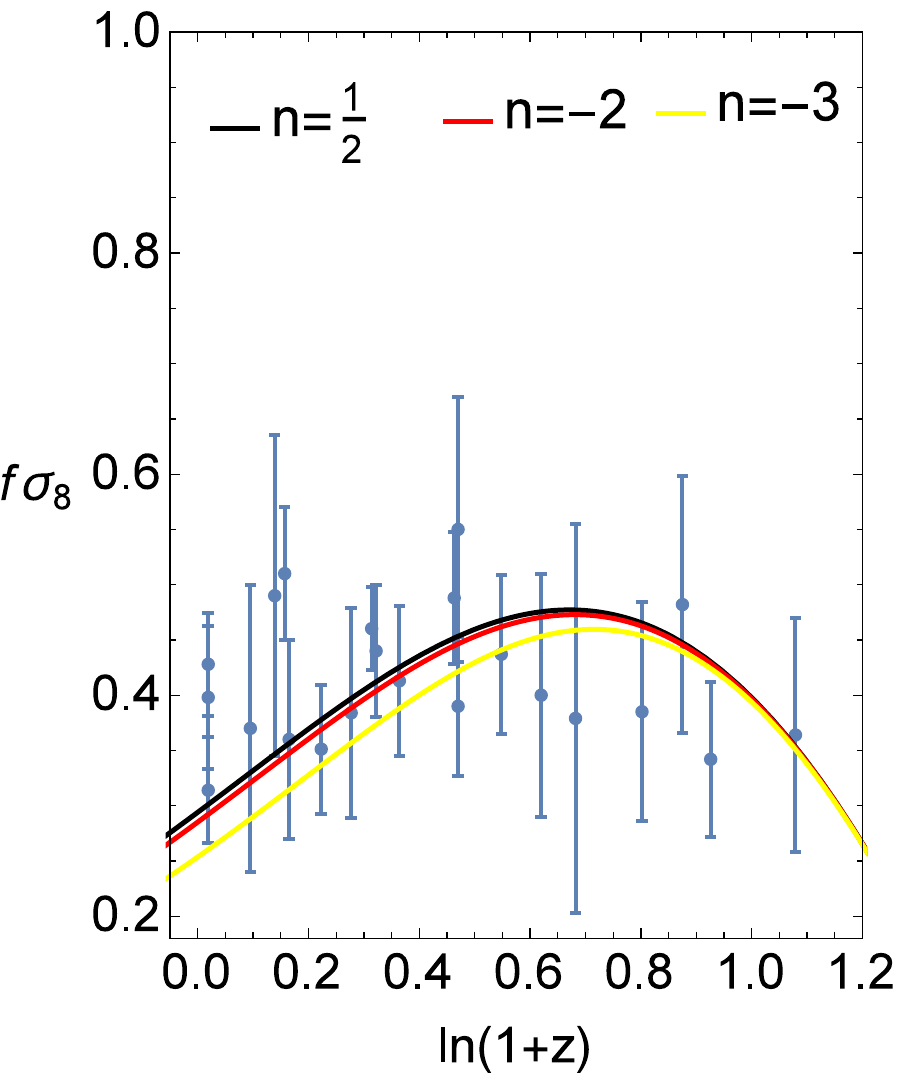} 
\label{fig:fig3_fsigma8_pow}
		\caption{{\it{The evolution of $f\sigma_8$ as a function of the 
redshift, for the power-law model I of (\ref{ModelI}), for various values of 
$n$. The data are taken from \cite{Sagredo:2018ahx,Kazantzidis:2018rnb}.}}}
	\end{figure}

 Finally, in order to test the predictions on the matter growth   with 
observational data, in Fig. 
\ref{fig:fig3_fsigma8_pow} we provide the evolution of     $f\sigma_8$. 
This quantity is defined as the product of the growth rate factor, $f= u= 
\frac{d \ln \delta}{d\ln a}$, and the root-mean-square normalization of the 
matter power spectrum  $\sigma_8$. The value of $\sigma_8$ usually depends on 
the model, however  here we have taken  $\sigma_8=0.8$  which could alleviate 
the present $\sigma_8$ tension between the RSD and Planck data 
\cite{Abdalla:2022yfr}. We mention here that we have checked that for $n>0$  
the evolution of $f\sigma_8$ coincides with that of   $\Lambda$CDM 
scenario, namely with the case $n=0$. From  Fig. \ref{fig:fig3_fsigma8_pow},  we 
observe that models with $n<0$ have smaller value of $f\sigma_8$, however, the 
data   prefer comparatively larger values. Hence, models with $n<0$ are 
not favored by the data. Additionally, it is worth noting that 
observational data still favour $n<1$ 
\cite{Lazkoz:2019sjl,Barros:2020bgg,Atayde:2021pgb,Ayuso:2020dcu}. Thus, from 
our analysis  we can conclude that the condition $0<n<1$ is required in order 
to acquire   consistency with observations.
	
	\subsection{ Model II: $F(Q)=Q 
e^{\beta\frac{Q_0}{Q}}-Q$}\label{sec:exp_model}

	In this subsection we consider  the exponential model 
\cite{Anagnostopoulos:2021ydo}
	 	\begin{eqnarray}
	 	\label{ModelII}
F(Q)=Q 
e^{\beta\frac{Q_0}{Q}}-Q,
 	\end{eqnarray} 
	  with $\beta$ the only dimensionless parameter. For $\beta=0$  the 
model is equivalent to GR without a cosmological constant, however the 
interesting feature is that for $\beta\neq0$ this model can fit observations in 
a very satisfactory way, although it does not include  a cosmological constant 
\cite{Anagnostopoulos:2021ydo}. Note that applying the first Friedmann equation 
at 
present, $\beta$ can be eliminated in terms of
$\Omega_{m0}$ \cite{Anagnostopoulos:2021ydo}. Additionally, since at early times 
$Q\gg Q_0$, 
the model tends to GR limit   and hence it trivially passes the 
BBN constraints  
\cite{Anagnostopoulos:2021ydo,Anagnostopoulos:2022gej}.  
	
In this case  we have $QF_{QQ}=\frac {\left( x+1 \right) 
^{2}+x \left( y-2 \right) +\frac{y^2}{4}-1}{x+1}$, and therefore the dynamical 
system \eqref{eq:xp}-\eqref{eq:up}  has four curves of critical points. In what 
follows,  we shall describe the properties of each curve.

	\begin{itemize}
		\item {\it  Curve of critical points $A_2$ $\left(-\frac{y}{2}, 
y,1\right)$}: This curve corresponds to a matter scaling solution with 
$\Omega_{m}=1-\frac{y}{2}$ and $w_{\rm eff}=0$. The corresponding Jacobian 
matrix  has  the eigenvalues $-\frac{5}{2}, 3$ and $0$, and thus  the 
corresponding points are always  saddle. Furthermore, since $u=1$, 
it is implied that the matter perturbations grow, and hence  the solution is of 
interest from the structure formation point of view. 
		
 \item {\it Curve of critical points  $B_2$ $\left(-\frac{y}{2}, 
y,-\frac{3}{2}\right)$}: Similarly to   $A_2$, this curve corresponds to a 
matter scaling solution. However, since $u<0$, we deduce that the matter 
perturbations decay and  therefore it cannot   describe the growth of 
structures during the matter-dominated epoch. It corresponds to an unstable node 
with eigenvalues $\frac{5}{2}, 3$ and $0$.
		
				\begin{figure}[ht]
	\centering
	\includegraphics[width=7cm, height=7.2cm]{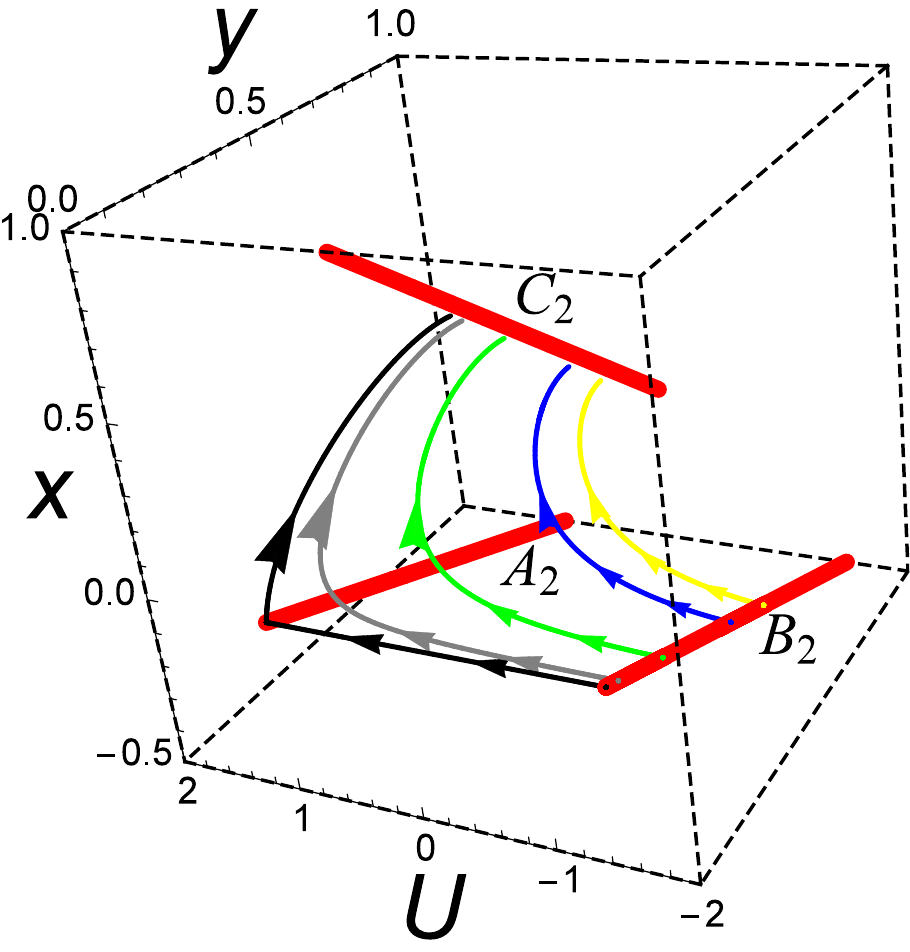} 
\label{fig:fig3_expo_phase}
	\caption{{\it{The phase portrait of the system 
\eqref{eq:xp}-\eqref{eq:up}, for the  exponential model II of (\ref{ModelII}). 
This particular example exhibits  the evolution  $B_2 
\to A_2\to C_2$.}} }
\end{figure}

 \item {\it Curves of critical points  $C_2$ $\left(1-y, y,0\right)$ and 
$D_2$ $\left(1-y, y,-2\right)$}: Both these curves correspond to   
accelerated solutions, dominated by the non-metricity component. The 
stability and cosmological properties of curves $C_2$ and  $D_2$ are exactly the
same with the curves  $C_1$ and $D_1$. Finally, we   find that only   
  $C_2$ is interesting for the late-time universe at the perturbation level.  
	\end{itemize}
 `

\begin{figure}[ht]
	\centering
		\includegraphics[width=6.cm, height=6.cm]{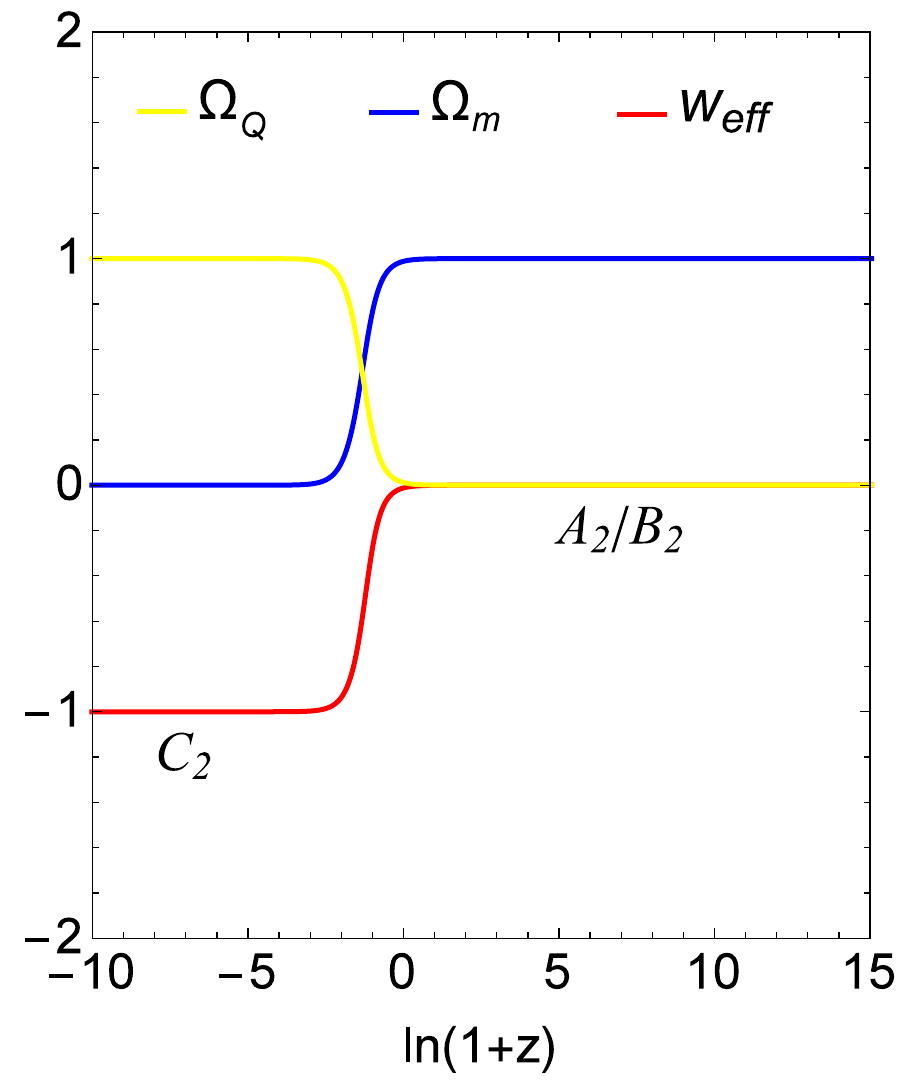} 
\label{fig:fig4_expo_para}
		\includegraphics[width=6.cm, height=6.cm]{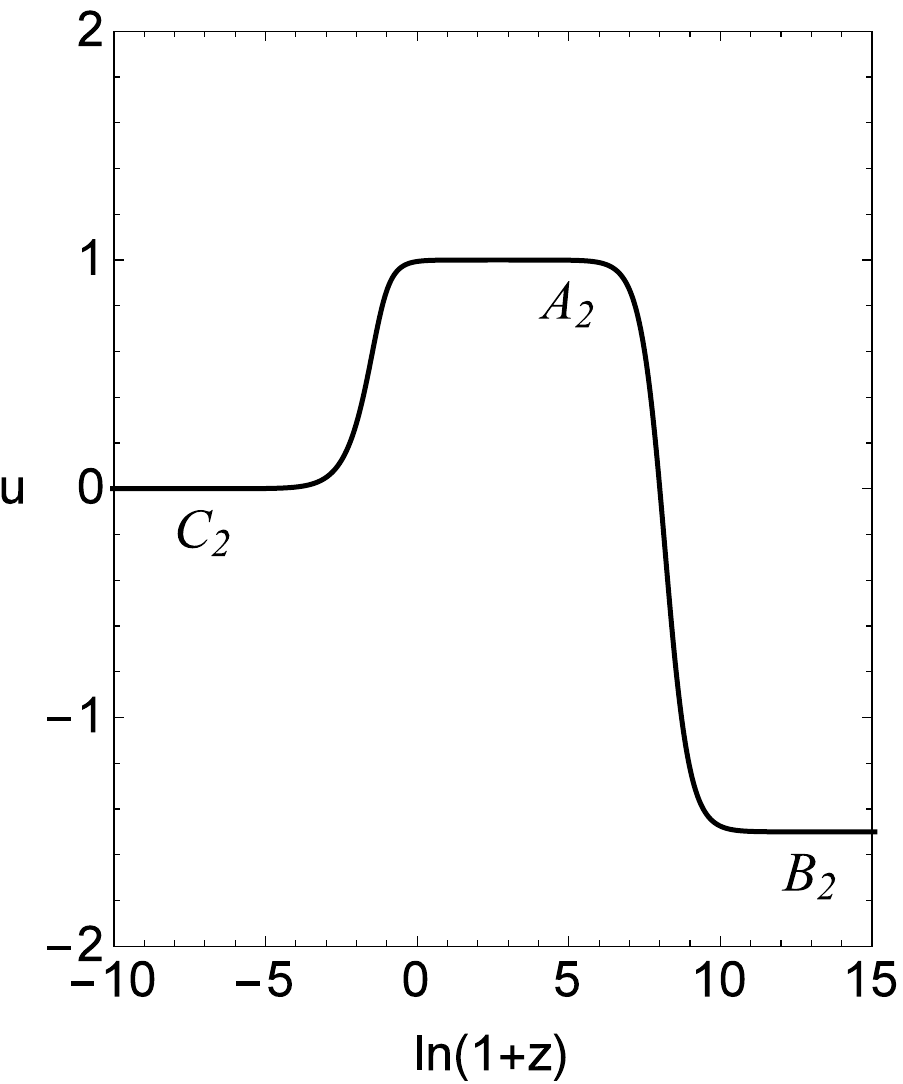} 
	\caption{{\it{ Upper graph:   Evolution of the density parameters of matter 
$\Omega_{m}$  and of  non-metricity (dark-energy) $\Omega_{Q}$, 
 as well as of the total, effective 
equation-of-state parameter $w_{\rm eff}$, as functions of the redshift, for   
the  exponential model II of (\ref{ModelII}).   Lower graph:   
Evolution     of the perturbation variable $u=\frac{d(\ln \delta)}{d(\ln a)}$. 
}} }
	\label{fig:fig4_expo}
\end{figure}

Similarly to Model I of the previous subsection, we see that 
 the inclusion of perturbations distinguishes   critical points that are 
equivalent at the background	level. Hence, from the combined background and 
perturbation analysis  we find that only curve $A_2$ is physically interesting 
to describe the matter-dominated epoch, where matter perturbations are 
generated. On the other hand, curve $C_2$ corresponds to late-time dark-energy 
domination, with a fixed evolution of matter perturbations, as observations 
require.  Lastly, for this model we also  find   no critical points at infinity.

In Fig. 
\ref{fig:fig3_expo_phase} we present the phase-space evolution, describing the
transition $B_2 \to A_2 \to C_2$.  Furthermore, in Fig. \ref{fig:fig4_expo}  we
	depict the evolution of the background cosmological parameters and the 
growth rate $u$, where we observe   the transition from matter domination 
towards a late-time dark-energy  dominated epoch.  	
As we mentioned above, it is interesting that even though the present 
model does not possess a $\Lambda$CDM limit for any parameter choice, the 
corresponding dynamics   is   qualitatively similar  with that of  
$\Lambda$CDM (see Fig. \ref{fig:fig4_expo}). Hence, since the model is free 
from the cosmological constant problem, it may be considered as   
slightly preferred over the $\Lambda$CDM scenario,  constituting  an interesting
alternative.

\begin{figure}[h!]
		\centering
		\includegraphics[width=8cm, height=7.2cm]{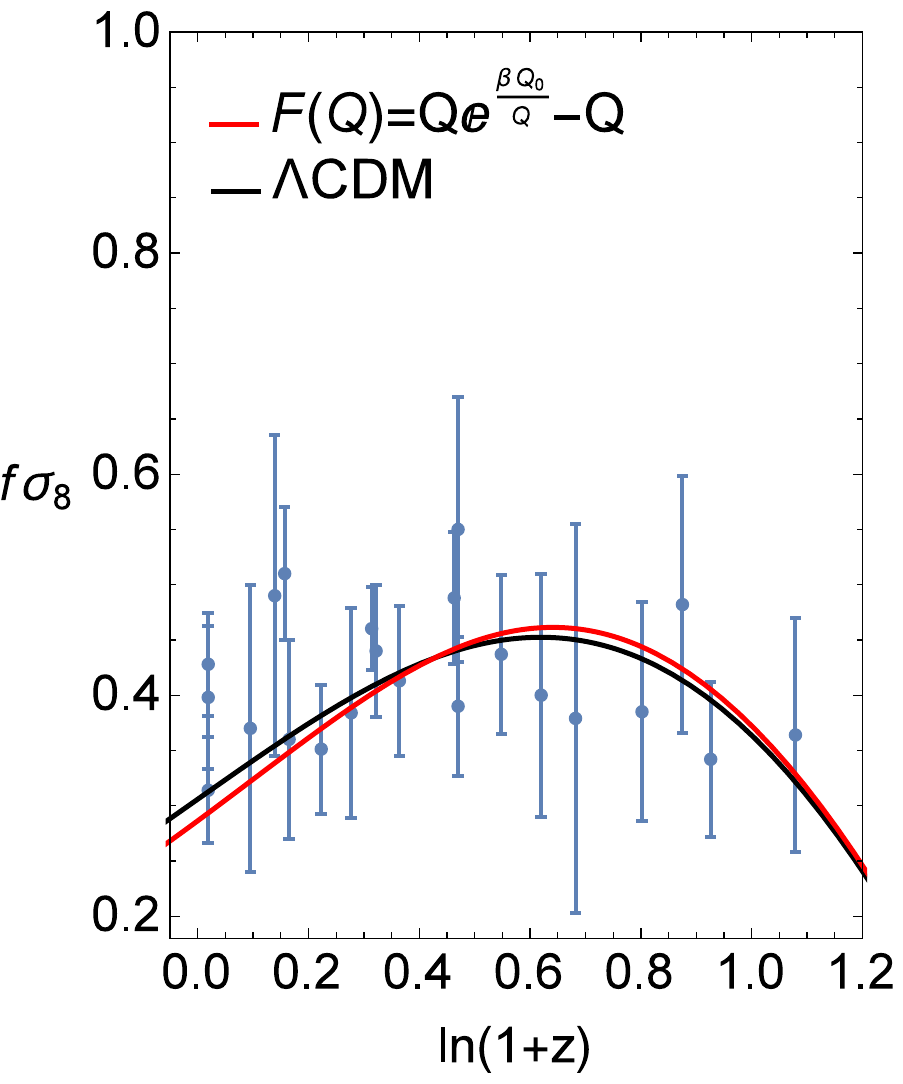} \label{fig:fig6_fsigma8_exp}
		\caption{
		{\it{The evolution of $f\sigma_8$ as a function of the 
redshift, for the  exponential model II of (\ref{ModelII}).   
The data are taken from \cite{Sagredo:2018ahx,Kazantzidis:2018rnb}.}}}
	\end{figure}

Finally,  in   Fig. \ref{fig:fig6_fsigma8_exp}    we investigate the 
evolution of $f\sigma_8$, using      
$\sigma_8=0.7$. As we observe, the behavior is 
comparable with that of    $\Lambda$CDM  paradigm. In summary,  the unified 
dynamical system analysis confirms the observational investigation at background 
and perturbation levels performed in 
\cite{Anagnostopoulos:2021ydo,Atayde:2021pgb}.
	
	\section{Conclusions}\label{sec:conc}

 Motivated by the fact that cosmological models based on $f(Q)$ 
 gravity are very efficient in fitting observational datasets at both 
 background and perturbation levels \cite{Anagnostopoulos:2021ydo,Atayde:2021pgb}, in the present work we 
 performed a  combined dynamical system analysis of both background and perturbation equations in order to examine the validity of this result through an independent method. 
 
 After transforming the background and perturbation equations into an autonomous system, we focused on two studied $f(Q)$ models of the literature, namely the power-law and the exponential ones. Due to the extra variable related to matter perturbations, each background critical point split into two points, characterized by different behavior of matter perturbations and stability.
 
 Concerning the power-law model, we obtained a matter-dominated saddle point characterized by the correct growth rate of matter perturbations, followed by the transition to a stable dark-energy dominated accelerated universe in which matter perturbations remain constant. Furthermore, we studied the growth of matter perturbations   by  analyzing the behavior of $f \sigma_8$, and we  found that the   model fits observational data successfully if the exponent lies within $0<n<1$, in which case we obtained a behavior similar to that of 
 $\Lambda$CDM scenario.
 
 Concerning the exponential model, we also found  curves of  points 
 corresponding to matter domination and matter perturbation growth,  and the fact that they are saddle provides a successful transition to the stable late-time dark-energy dominated solution with constant matter perturbations. The interesting feature of this model is that this desired behavior is obtained although the model does not possess $\Lambda$CDM scenario as a particular limit, namely, it arises solely from the non-metricity structure. Additionally, we found that 
 while the power-law model resembles   $\Lambda$CDM cosmology for $n<1$, the exponential model resembles the latter for any choice of the model parameter. 
 
 In summary,  the combined dynamical analysis at the  background and 
 perturbation levels do verify the  results of observational confrontation, showing through an independent way that $f(Q)$ gravity, and 
 specifically, the exponential model can be considered as a very promising 
 alternative to  $\Lambda$CDM concordance model.

	\begin{acknowledgments}
		JD was supported by the Core Research Grant of SERB, Department of 
Science and Technology India (File No. CRG $\slash 2018 \slash 001035$) and the 
Associate program of IUCAA.  This work is partially supported by the Ministry of 
Education and Science of the Republic of Kazakhstan, Grant AP08856912.  ENS 
also acknowledge participation in the COST Association 
Action CA18108 ``{\it Quantum Gravity Phenomenology in the Multimessenger 
Approach (QG-MM)}''.
	\end{acknowledgments}

	\bibliography{fqrefs}

\begin{thebibliography}{90}%
\makeatletter
\providecommand \@ifxundefined [1]{%
 \@ifx{#1\undefined}
}%
\providecommand \@ifnum [1]{%
 \ifnum #1\expandafter \@firstoftwo
 \else \expandafter \@secondoftwo
 \fi
}%
\providecommand \@ifx [1]{%
 \ifx #1\expandafter \@firstoftwo
 \else \expandafter \@secondoftwo
 \fi
}%
\providecommand \natexlab [1]{#1}%
\providecommand \enquote  [1]{``#1''}%
\providecommand \bibnamefont  [1]{#1}%
\providecommand \bibfnamefont [1]{#1}%
\providecommand \citenamefont [1]{#1}%
\providecommand \href@noop [0]{\@secondoftwo}%
\providecommand \href [0]{\begingroup \@sanitize@url \@href}%
\providecommand \@href[1]{\@@startlink{#1}\@@href}%
\providecommand \@@href[1]{\endgroup#1\@@endlink}%
\providecommand \@sanitize@url [0]{\catcode `\\12\catcode `\$12\catcode
  `\&12\catcode `\#12\catcode `\^12\catcode `\_12\catcode `\%12\relax}%
\providecommand \@@startlink[1]{}%
\providecommand \@@endlink[0]{}%
\providecommand \url  [0]{\begingroup\@sanitize@url \@url }%
\providecommand \@url [1]{\endgroup\@href {#1}{\urlprefix }}%
\providecommand \urlprefix  [0]{URL }%
\providecommand \Eprint [0]{\href }%
\providecommand \doibase [0]{http://dx.doi.org/}%
\providecommand \selectlanguage [0]{\@gobble}%
\providecommand \bibinfo  [0]{\@secondoftwo}%
\providecommand \bibfield  [0]{\@secondoftwo}%
\providecommand \translation [1]{[#1]}%
\providecommand \BibitemOpen [0]{}%
\providecommand \bibitemStop [0]{}%
\providecommand \bibitemNoStop [0]{.\EOS\space}%
\providecommand \EOS [0]{\spacefactor3000\relax}%
\providecommand \BibitemShut  [1]{\csname bibitem#1\endcsname}%
\let\auto@bib@innerbib\@empty
\bibitem [{\citenamefont {Addazi}\ \emph {et~al.}(2022)\citenamefont {Addazi}
  \emph {et~al.}}]{Addazi:2021xuf}%
  \BibitemOpen
  \bibfield  {author} {\bibinfo {author} {\bibfnamefont {A.}~\bibnamefont
  {Addazi}} \emph {et~al.},\ }\bibfield  {title} {\enquote {\bibinfo {title}
  {{Quantum gravity phenomenology at the dawn of the multi-messenger
  era\textemdash{}A review}},}\ }\href {\doibase 10.1016/j.ppnp.2022.103948}
  {\bibfield  {journal} {\bibinfo  {journal} {Prog. Part. Nucl. Phys.}\
  }\textbf {\bibinfo {volume} {125}},\ \bibinfo {pages} {103948} (\bibinfo
  {year} {2022})},\ \Eprint {http://arxiv.org/abs/2111.05659} {arXiv:2111.05659
  [hep-ph]} \BibitemShut {NoStop}%
\bibitem [{\citenamefont {Abdalla}\ \emph {et~al.}(2022)\citenamefont {Abdalla}
  \emph {et~al.}}]{Abdalla:2022yfr}%
  \BibitemOpen
  \bibfield  {author} {\bibinfo {author} {\bibfnamefont {Elcio}\ \bibnamefont
  {Abdalla}} \emph {et~al.},\ }\bibfield  {title} {\enquote {\bibinfo {title}
  {{Cosmology intertwined: A review of the particle physics, astrophysics, and
  cosmology associated with the cosmological tensions and anomalies}},}\ }\href
  {\doibase 10.1016/j.jheap.2022.04.002} {\bibfield  {journal} {\bibinfo
  {journal} {JHEAp}\ }\textbf {\bibinfo {volume} {34}},\ \bibinfo {pages}
  {49--211} (\bibinfo {year} {2022})},\ \Eprint
  {http://arxiv.org/abs/2203.06142} {arXiv:2203.06142 [astro-ph.CO]}
  \BibitemShut {NoStop}%
\bibitem [{\citenamefont {Martin}(2012)}]{Martin:2012bt}%
  \BibitemOpen
  \bibfield  {author} {\bibinfo {author} {\bibfnamefont {Jerome}\ \bibnamefont
  {Martin}},\ }\bibfield  {title} {\enquote {\bibinfo {title} {{Everything You
  Always Wanted To Know About The Cosmological Constant Problem (But Were
  Afraid To Ask)}},}\ }\href {\doibase 10.1016/j.crhy.2012.04.008} {\bibfield
  {journal} {\bibinfo  {journal} {Comptes Rendus Physique}\ }\textbf {\bibinfo
  {volume} {13}},\ \bibinfo {pages} {566--665} (\bibinfo {year} {2012})},\
  \Eprint {http://arxiv.org/abs/1205.3365} {arXiv:1205.3365 [astro-ph.CO]}
  \BibitemShut {NoStop}%
\bibitem [{\citenamefont {Freedman}(2017)}]{Freedman:2017yms}%
  \BibitemOpen
  \bibfield  {author} {\bibinfo {author} {\bibfnamefont {Wendy~L.}\
  \bibnamefont {Freedman}},\ }\bibfield  {title} {\enquote {\bibinfo {title}
  {{Cosmology at a Crossroads}},}\ }\href {\doibase 10.1038/s41550-017-0121}
  {\bibfield  {journal} {\bibinfo  {journal} {Nature Astron.}\ }\textbf
  {\bibinfo {volume} {1}},\ \bibinfo {pages} {0121} (\bibinfo {year} {2017})},\
  \Eprint {http://arxiv.org/abs/1706.02739} {arXiv:1706.02739 [astro-ph.CO]}
  \BibitemShut {NoStop}%
\bibitem [{\citenamefont {Lusso}\ \emph {et~al.}(2019)\citenamefont {Lusso},
  \citenamefont {Piedipalumbo}, \citenamefont {Risaliti}, \citenamefont
  {Paolillo}, \citenamefont {Bisogni}, \citenamefont {Nardini},\ and\
  \citenamefont {Amati}}]{Lusso:2019akb}%
  \BibitemOpen
  \bibfield  {author} {\bibinfo {author} {\bibfnamefont {E.}~\bibnamefont
  {Lusso}}, \bibinfo {author} {\bibfnamefont {E.}~\bibnamefont {Piedipalumbo}},
  \bibinfo {author} {\bibfnamefont {G.}~\bibnamefont {Risaliti}}, \bibinfo
  {author} {\bibfnamefont {M.}~\bibnamefont {Paolillo}}, \bibinfo {author}
  {\bibfnamefont {S.}~\bibnamefont {Bisogni}}, \bibinfo {author} {\bibfnamefont
  {E.}~\bibnamefont {Nardini}}, \ and\ \bibinfo {author} {\bibfnamefont
  {L.}~\bibnamefont {Amati}},\ }\bibfield  {title} {\enquote {\bibinfo {title}
  {{Tension with the flat $\Lambda$CDM model from a high-redshift Hubble
  diagram of supernovae, quasars, and gamma-ray bursts}},}\ }\href {\doibase
  10.1051/0004-6361/201936223} {\bibfield  {journal} {\bibinfo  {journal}
  {Astron. Astrophys.}\ }\textbf {\bibinfo {volume} {628}},\ \bibinfo {pages}
  {L4} (\bibinfo {year} {2019})},\ \Eprint {http://arxiv.org/abs/1907.07692}
  {arXiv:1907.07692 [astro-ph.CO]} \BibitemShut {NoStop}%
\bibitem [{\citenamefont {Lin}\ \emph {et~al.}(2020)\citenamefont {Lin},
  \citenamefont {Mack},\ and\ \citenamefont {Hou}}]{Lin:2019htv}%
  \BibitemOpen
  \bibfield  {author} {\bibinfo {author} {\bibfnamefont {Weikang}\ \bibnamefont
  {Lin}}, \bibinfo {author} {\bibfnamefont {Katherine~J.}\ \bibnamefont
  {Mack}}, \ and\ \bibinfo {author} {\bibfnamefont {Liqiang}\ \bibnamefont
  {Hou}},\ }\bibfield  {title} {\enquote {\bibinfo {title} {{Investigating the
  Hubble Constant Tension -- Two Numbers in the Standard Cosmological
  Model}},}\ }\href {\doibase 10.3847/2041-8213/abc894} {\bibfield  {journal}
  {\bibinfo  {journal} {Astrophys. J. Lett.}\ }\textbf {\bibinfo {volume}
  {904}},\ \bibinfo {pages} {L22} (\bibinfo {year} {2020})},\ \Eprint
  {http://arxiv.org/abs/1910.02978} {arXiv:1910.02978 [astro-ph.CO]}
  \BibitemShut {NoStop}%
\bibitem [{\citenamefont {Perivolaropoulos}\ and\ \citenamefont
  {Skara}(2021)}]{Perivolaropoulos:2021jda}%
  \BibitemOpen
  \bibfield  {author} {\bibinfo {author} {\bibfnamefont {Leandros}\
  \bibnamefont {Perivolaropoulos}}\ and\ \bibinfo {author} {\bibfnamefont
  {Foteini}\ \bibnamefont {Skara}},\ }\bibfield  {title} {\enquote {\bibinfo
  {title} {{Challenges for $\Lambda$CDM: An update}},}\ }\href@noop {} {\
  (\bibinfo {year} {2021})},\ \Eprint {http://arxiv.org/abs/2105.05208}
  {arXiv:2105.05208 [astro-ph.CO]} \BibitemShut {NoStop}%
\bibitem [{\citenamefont {Saridakis}\ \emph {et~al.}(2021)\citenamefont
  {Saridakis} \emph {et~al.}}]{CANTATA:2021ktz}%
  \BibitemOpen
  \bibfield  {author} {\bibinfo {author} {\bibfnamefont {Emmanuel~N.}\
  \bibnamefont {Saridakis}} \emph {et~al.} (\bibinfo {collaboration}
  {CANTATA}),\ }\bibfield  {title} {\enquote {\bibinfo {title} {{Modified
  Gravity and Cosmology: An Update by the CANTATA Network}},}\ }\href@noop {}
  {\  (\bibinfo {year} {2021})},\ \Eprint {http://arxiv.org/abs/2105.12582}
  {arXiv:2105.12582 [gr-qc]} \BibitemShut {NoStop}%
\bibitem [{\citenamefont {Capozziello}\ and\ \citenamefont
  {De~Laurentis}(2011)}]{Capozziello:2011et}%
  \BibitemOpen
  \bibfield  {author} {\bibinfo {author} {\bibfnamefont {Salvatore}\
  \bibnamefont {Capozziello}}\ and\ \bibinfo {author} {\bibfnamefont
  {Mariafelicia}\ \bibnamefont {De~Laurentis}},\ }\bibfield  {title} {\enquote
  {\bibinfo {title} {{Extended Theories of Gravity}},}\ }\href {\doibase
  10.1016/j.physrep.2011.09.003} {\bibfield  {journal} {\bibinfo  {journal}
  {Phys. Rept.}\ }\textbf {\bibinfo {volume} {509}},\ \bibinfo {pages}
  {167--321} (\bibinfo {year} {2011})},\ \Eprint
  {http://arxiv.org/abs/1108.6266} {arXiv:1108.6266 [gr-qc]} \BibitemShut
  {NoStop}%
\bibitem [{\citenamefont {Clifton}\ \emph {et~al.}(2012)\citenamefont
  {Clifton}, \citenamefont {Ferreira}, \citenamefont {Padilla},\ and\
  \citenamefont {Skordis}}]{Clifton:2011jh}%
  \BibitemOpen
  \bibfield  {author} {\bibinfo {author} {\bibfnamefont {Timothy}\ \bibnamefont
  {Clifton}}, \bibinfo {author} {\bibfnamefont {Pedro~G.}\ \bibnamefont
  {Ferreira}}, \bibinfo {author} {\bibfnamefont {Antonio}\ \bibnamefont
  {Padilla}}, \ and\ \bibinfo {author} {\bibfnamefont {Constantinos}\
  \bibnamefont {Skordis}},\ }\bibfield  {title} {\enquote {\bibinfo {title}
  {{Modified Gravity and Cosmology}},}\ }\href {\doibase
  10.1016/j.physrep.2012.01.001} {\bibfield  {journal} {\bibinfo  {journal}
  {Phys. Rept.}\ }\textbf {\bibinfo {volume} {513}},\ \bibinfo {pages} {1--189}
  (\bibinfo {year} {2012})},\ \Eprint {http://arxiv.org/abs/1106.2476}
  {arXiv:1106.2476 [astro-ph.CO]} \BibitemShut {NoStop}%
\bibitem [{\citenamefont {Cai}\ \emph {et~al.}(2016)\citenamefont {Cai},
  \citenamefont {Capozziello}, \citenamefont {De~Laurentis},\ and\
  \citenamefont {Saridakis}}]{Cai:2015emx}%
  \BibitemOpen
  \bibfield  {author} {\bibinfo {author} {\bibfnamefont {Yi-Fu}\ \bibnamefont
  {Cai}}, \bibinfo {author} {\bibfnamefont {Salvatore}\ \bibnamefont
  {Capozziello}}, \bibinfo {author} {\bibfnamefont {Mariafelicia}\ \bibnamefont
  {De~Laurentis}}, \ and\ \bibinfo {author} {\bibfnamefont {Emmanuel~N.}\
  \bibnamefont {Saridakis}},\ }\bibfield  {title} {\enquote {\bibinfo {title}
  {{f(T) teleparallel gravity and cosmology}},}\ }\href {\doibase
  10.1088/0034-4885/79/10/106901} {\bibfield  {journal} {\bibinfo  {journal}
  {Rept. Prog. Phys.}\ }\textbf {\bibinfo {volume} {79}},\ \bibinfo {pages}
  {106901} (\bibinfo {year} {2016})},\ \Eprint
  {http://arxiv.org/abs/1511.07586} {arXiv:1511.07586 [gr-qc]} \BibitemShut
  {NoStop}%
\bibitem [{\citenamefont {Copeland}\ \emph {et~al.}(2006)\citenamefont
  {Copeland}, \citenamefont {Sami},\ and\ \citenamefont
  {Tsujikawa}}]{Copeland:2006wr}%
  \BibitemOpen
  \bibfield  {author} {\bibinfo {author} {\bibfnamefont {Edmund~J.}\
  \bibnamefont {Copeland}}, \bibinfo {author} {\bibfnamefont {M.}~\bibnamefont
  {Sami}}, \ and\ \bibinfo {author} {\bibfnamefont {Shinji}\ \bibnamefont
  {Tsujikawa}},\ }\bibfield  {title} {\enquote {\bibinfo {title} {{Dynamics of
  dark energy}},}\ }\href {\doibase 10.1142/S021827180600942X} {\bibfield
  {journal} {\bibinfo  {journal} {Int. J. Mod. Phys. D}\ }\textbf {\bibinfo
  {volume} {15}},\ \bibinfo {pages} {1753--1936} (\bibinfo {year} {2006})},\
  \Eprint {http://arxiv.org/abs/hep-th/0603057} {arXiv:hep-th/0603057}
  \BibitemShut {NoStop}%
\bibitem [{\citenamefont {Joyce}\ \emph {et~al.}(2016)\citenamefont {Joyce},
  \citenamefont {Lombriser},\ and\ \citenamefont {Schmidt}}]{Joyce:2016vqv}%
  \BibitemOpen
  \bibfield  {author} {\bibinfo {author} {\bibfnamefont {Austin}\ \bibnamefont
  {Joyce}}, \bibinfo {author} {\bibfnamefont {Lucas}\ \bibnamefont
  {Lombriser}}, \ and\ \bibinfo {author} {\bibfnamefont {Fabian}\ \bibnamefont
  {Schmidt}},\ }\bibfield  {title} {\enquote {\bibinfo {title} {{Dark Energy
  Versus Modified Gravity}},}\ }\href {\doibase
  10.1146/annurev-nucl-102115-044553} {\bibfield  {journal} {\bibinfo
  {journal} {Ann. Rev. Nucl. Part. Sci.}\ }\textbf {\bibinfo {volume} {66}},\
  \bibinfo {pages} {95--122} (\bibinfo {year} {2016})},\ \Eprint
  {http://arxiv.org/abs/1601.06133} {arXiv:1601.06133 [astro-ph.CO]}
  \BibitemShut {NoStop}%
\bibitem [{\citenamefont {Cai}\ \emph {et~al.}(2010)\citenamefont {Cai},
  \citenamefont {Saridakis}, \citenamefont {Setare},\ and\ \citenamefont
  {Xia}}]{Cai:2009zp}%
  \BibitemOpen
  \bibfield  {author} {\bibinfo {author} {\bibfnamefont {Yi-Fu}\ \bibnamefont
  {Cai}}, \bibinfo {author} {\bibfnamefont {Emmanuel~N.}\ \bibnamefont
  {Saridakis}}, \bibinfo {author} {\bibfnamefont {Mohammad~R.}\ \bibnamefont
  {Setare}}, \ and\ \bibinfo {author} {\bibfnamefont {Jun-Qing}\ \bibnamefont
  {Xia}},\ }\bibfield  {title} {\enquote {\bibinfo {title} {{Quintom Cosmology:
  Theoretical implications and observations}},}\ }\href {\doibase
  10.1016/j.physrep.2010.04.001} {\bibfield  {journal} {\bibinfo  {journal}
  {Phys. Rept.}\ }\textbf {\bibinfo {volume} {493}},\ \bibinfo {pages} {1--60}
  (\bibinfo {year} {2010})},\ \Eprint {http://arxiv.org/abs/0909.2776}
  {arXiv:0909.2776 [hep-th]} \BibitemShut {NoStop}%
\bibitem [{\citenamefont {Starobinsky}(1980)}]{Starobinsky:1980te}%
  \BibitemOpen
  \bibfield  {author} {\bibinfo {author} {\bibfnamefont {Alexei~A.}\
  \bibnamefont {Starobinsky}},\ }\bibfield  {title} {\enquote {\bibinfo {title}
  {{A New Type of Isotropic Cosmological Models Without Singularity}},}\ }\href
  {\doibase 10.1016/0370-2693(80)90670-X} {\bibfield  {journal} {\bibinfo
  {journal} {Phys. Lett. B}\ }\textbf {\bibinfo {volume} {91}},\ \bibinfo
  {pages} {99--102} (\bibinfo {year} {1980})}\BibitemShut {NoStop}%
\bibitem [{\citenamefont {Capozziello}(2002)}]{Capozziello:2002rd}%
  \BibitemOpen
  \bibfield  {author} {\bibinfo {author} {\bibfnamefont {Salvatore}\
  \bibnamefont {Capozziello}},\ }\bibfield  {title} {\enquote {\bibinfo {title}
  {{Curvature quintessence}},}\ }\href {\doibase 10.1142/S0218271802002025}
  {\bibfield  {journal} {\bibinfo  {journal} {Int. J. Mod. Phys. D}\ }\textbf
  {\bibinfo {volume} {11}},\ \bibinfo {pages} {483--492} (\bibinfo {year}
  {2002})},\ \Eprint {http://arxiv.org/abs/gr-qc/0201033} {arXiv:gr-qc/0201033}
  \BibitemShut {NoStop}%
\bibitem [{\citenamefont {De~Felice}\ and\ \citenamefont
  {Tsujikawa}(2010)}]{DeFelice:2010aj}%
  \BibitemOpen
  \bibfield  {author} {\bibinfo {author} {\bibfnamefont {Antonio}\ \bibnamefont
  {De~Felice}}\ and\ \bibinfo {author} {\bibfnamefont {Shinji}\ \bibnamefont
  {Tsujikawa}},\ }\bibfield  {title} {\enquote {\bibinfo {title} {{f(R)
  theories}},}\ }\href {\doibase 10.12942/lrr-2010-3} {\bibfield  {journal}
  {\bibinfo  {journal} {Living Rev. Rel.}\ }\textbf {\bibinfo {volume} {13}},\
  \bibinfo {pages} {3} (\bibinfo {year} {2010})},\ \Eprint
  {http://arxiv.org/abs/1002.4928} {arXiv:1002.4928 [gr-qc]} \BibitemShut
  {NoStop}%
\bibitem [{\citenamefont {Antoniadis}\ \emph {et~al.}(1994)\citenamefont
  {Antoniadis}, \citenamefont {Rizos},\ and\ \citenamefont
  {Tamvakis}}]{Antoniadis:1993jc}%
  \BibitemOpen
  \bibfield  {author} {\bibinfo {author} {\bibfnamefont {Ignatios}\
  \bibnamefont {Antoniadis}}, \bibinfo {author} {\bibfnamefont
  {J.}~\bibnamefont {Rizos}}, \ and\ \bibinfo {author} {\bibfnamefont
  {K.}~\bibnamefont {Tamvakis}},\ }\bibfield  {title} {\enquote {\bibinfo
  {title} {{Singularity - free cosmological solutions of the superstring
  effective action}},}\ }\href {\doibase 10.1016/0550-3213(94)90120-1}
  {\bibfield  {journal} {\bibinfo  {journal} {Nucl. Phys. B}\ }\textbf
  {\bibinfo {volume} {415}},\ \bibinfo {pages} {497--514} (\bibinfo {year}
  {1994})},\ \Eprint {http://arxiv.org/abs/hep-th/9305025}
  {arXiv:hep-th/9305025} \BibitemShut {NoStop}%
\bibitem [{\citenamefont {Nojiri}\ and\ \citenamefont
  {Odintsov}(2005)}]{Nojiri:2005jg}%
  \BibitemOpen
  \bibfield  {author} {\bibinfo {author} {\bibfnamefont {Shin'ichi}\
  \bibnamefont {Nojiri}}\ and\ \bibinfo {author} {\bibfnamefont {Sergei~D.}\
  \bibnamefont {Odintsov}},\ }\bibfield  {title} {\enquote {\bibinfo {title}
  {{Modified Gauss-Bonnet theory as gravitational alternative for dark
  energy}},}\ }\href {\doibase 10.1016/j.physletb.2005.10.010} {\bibfield
  {journal} {\bibinfo  {journal} {Phys. Lett. B}\ }\textbf {\bibinfo {volume}
  {631}},\ \bibinfo {pages} {1--6} (\bibinfo {year} {2005})},\ \Eprint
  {http://arxiv.org/abs/hep-th/0508049} {arXiv:hep-th/0508049} \BibitemShut
  {NoStop}%
\bibitem [{\citenamefont {De~Felice}\ and\ \citenamefont
  {Tsujikawa}(2009)}]{DeFelice:2008wz}%
  \BibitemOpen
  \bibfield  {author} {\bibinfo {author} {\bibfnamefont {Antonio}\ \bibnamefont
  {De~Felice}}\ and\ \bibinfo {author} {\bibfnamefont {Shinji}\ \bibnamefont
  {Tsujikawa}},\ }\bibfield  {title} {\enquote {\bibinfo {title} {{Construction
  of cosmologically viable f(G) dark energy models}},}\ }\href {\doibase
  10.1016/j.physletb.2009.03.060} {\bibfield  {journal} {\bibinfo  {journal}
  {Phys. Lett. B}\ }\textbf {\bibinfo {volume} {675}},\ \bibinfo {pages} {1--8}
  (\bibinfo {year} {2009})},\ \Eprint {http://arxiv.org/abs/0810.5712}
  {arXiv:0810.5712 [hep-th]} \BibitemShut {NoStop}%
\bibitem [{\citenamefont {Erices}\ \emph {et~al.}(2019)\citenamefont {Erices},
  \citenamefont {Papantonopoulos},\ and\ \citenamefont
  {Saridakis}}]{Erices:2019mkd}%
  \BibitemOpen
  \bibfield  {author} {\bibinfo {author} {\bibfnamefont {Cristian}\
  \bibnamefont {Erices}}, \bibinfo {author} {\bibfnamefont {Eleftherios}\
  \bibnamefont {Papantonopoulos}}, \ and\ \bibinfo {author} {\bibfnamefont
  {Emmanuel~N.}\ \bibnamefont {Saridakis}},\ }\bibfield  {title} {\enquote
  {\bibinfo {title} {{Cosmology in cubic and $f(P)$ gravity}},}\ }\href
  {\doibase 10.1103/PhysRevD.99.123527} {\bibfield  {journal} {\bibinfo
  {journal} {Phys. Rev. D}\ }\textbf {\bibinfo {volume} {99}},\ \bibinfo
  {pages} {123527} (\bibinfo {year} {2019})},\ \Eprint
  {http://arxiv.org/abs/1903.11128} {arXiv:1903.11128 [gr-qc]} \BibitemShut
  {NoStop}%
\bibitem [{\citenamefont {Marciu}(2020)}]{Marciu:2020ysf}%
  \BibitemOpen
  \bibfield  {author} {\bibinfo {author} {\bibfnamefont {Mihai}\ \bibnamefont
  {Marciu}},\ }\bibfield  {title} {\enquote {\bibinfo {title} {{Note on the
  dynamical features for the extended $f(P)$ cubic gravity}},}\ }\href
  {\doibase 10.1103/PhysRevD.101.103534} {\bibfield  {journal} {\bibinfo
  {journal} {Phys. Rev. D}\ }\textbf {\bibinfo {volume} {101}},\ \bibinfo
  {pages} {103534} (\bibinfo {year} {2020})},\ \Eprint
  {http://arxiv.org/abs/2003.06403} {arXiv:2003.06403 [gr-qc]} \BibitemShut
  {NoStop}%
\bibitem [{\citenamefont {Beltr\'an~Jim\'enez}\ and\ \citenamefont
  {Jim\'enez-Cano}(2021)}]{BeltranJimenez:2020lee}%
  \BibitemOpen
  \bibfield  {author} {\bibinfo {author} {\bibfnamefont {Jose}\ \bibnamefont
  {Beltr\'an~Jim\'enez}}\ and\ \bibinfo {author} {\bibfnamefont {Alejandro}\
  \bibnamefont {Jim\'enez-Cano}},\ }\bibfield  {title} {\enquote {\bibinfo
  {title} {{On the strong coupling of Einsteinian Cubic Gravity and its
  generalisations}},}\ }\href {\doibase 10.1088/1475-7516/2021/01/069}
  {\bibfield  {journal} {\bibinfo  {journal} {JCAP}\ }\textbf {\bibinfo
  {volume} {01}},\ \bibinfo {pages} {069} (\bibinfo {year} {2021})},\ \Eprint
  {http://arxiv.org/abs/2009.08197} {arXiv:2009.08197 [gr-qc]} \BibitemShut
  {NoStop}%
\bibitem [{\citenamefont {Horndeski}(1974)}]{Horndeski:1974wa}%
  \BibitemOpen
  \bibfield  {author} {\bibinfo {author} {\bibfnamefont {Gregory~Walter}\
  \bibnamefont {Horndeski}},\ }\bibfield  {title} {\enquote {\bibinfo {title}
  {{Second-order scalar-tensor field equations in a four-dimensional space}},}\
  }\href {\doibase 10.1007/BF01807638} {\bibfield  {journal} {\bibinfo
  {journal} {Int. J. Theor. Phys.}\ }\textbf {\bibinfo {volume} {10}},\
  \bibinfo {pages} {363--384} (\bibinfo {year} {1974})}\BibitemShut {NoStop}%
\bibitem [{\citenamefont {Deffayet}\ \emph {et~al.}(2009)\citenamefont
  {Deffayet}, \citenamefont {Esposito-Farese},\ and\ \citenamefont
  {Vikman}}]{Deffayet:2009wt}%
  \BibitemOpen
  \bibfield  {author} {\bibinfo {author} {\bibfnamefont {C.}~\bibnamefont
  {Deffayet}}, \bibinfo {author} {\bibfnamefont {Gilles}\ \bibnamefont
  {Esposito-Farese}}, \ and\ \bibinfo {author} {\bibfnamefont {A.}~\bibnamefont
  {Vikman}},\ }\bibfield  {title} {\enquote {\bibinfo {title} {{Covariant
  Galileon}},}\ }\href {\doibase 10.1103/PhysRevD.79.084003} {\bibfield
  {journal} {\bibinfo  {journal} {Phys. Rev. D}\ }\textbf {\bibinfo {volume}
  {79}},\ \bibinfo {pages} {084003} (\bibinfo {year} {2009})},\ \Eprint
  {http://arxiv.org/abs/0901.1314} {arXiv:0901.1314 [hep-th]} \BibitemShut
  {NoStop}%
\bibitem [{\citenamefont {Bengochea}\ and\ \citenamefont
  {Ferraro}(2009)}]{Bengochea:2008gz}%
  \BibitemOpen
  \bibfield  {author} {\bibinfo {author} {\bibfnamefont {Gabriel~R.}\
  \bibnamefont {Bengochea}}\ and\ \bibinfo {author} {\bibfnamefont {Rafael}\
  \bibnamefont {Ferraro}},\ }\bibfield  {title} {\enquote {\bibinfo {title}
  {{Dark torsion as the cosmic speed-up}},}\ }\href {\doibase
  10.1103/PhysRevD.79.124019} {\bibfield  {journal} {\bibinfo  {journal} {Phys.
  Rev. D}\ }\textbf {\bibinfo {volume} {79}},\ \bibinfo {pages} {124019}
  (\bibinfo {year} {2009})},\ \Eprint {http://arxiv.org/abs/0812.1205}
  {arXiv:0812.1205 [astro-ph]} \BibitemShut {NoStop}%
\bibitem [{\citenamefont {Kofinas}\ and\ \citenamefont
  {Saridakis}(2014{\natexlab{a}})}]{Kofinas:2014owa}%
  \BibitemOpen
  \bibfield  {author} {\bibinfo {author} {\bibfnamefont {Georgios}\
  \bibnamefont {Kofinas}}\ and\ \bibinfo {author} {\bibfnamefont {Emmanuel~N.}\
  \bibnamefont {Saridakis}},\ }\bibfield  {title} {\enquote {\bibinfo {title}
  {{Teleparallel equivalent of Gauss-Bonnet gravity and its modifications}},}\
  }\href {\doibase 10.1103/PhysRevD.90.084044} {\bibfield  {journal} {\bibinfo
  {journal} {Phys. Rev. D}\ }\textbf {\bibinfo {volume} {90}},\ \bibinfo
  {pages} {084044} (\bibinfo {year} {2014}{\natexlab{a}})},\ \Eprint
  {http://arxiv.org/abs/1404.2249} {arXiv:1404.2249 [gr-qc]} \BibitemShut
  {NoStop}%
\bibitem [{\citenamefont {Kofinas}\ \emph {et~al.}(2014)\citenamefont
  {Kofinas}, \citenamefont {Leon},\ and\ \citenamefont
  {Saridakis}}]{Kofinas:2014aka}%
  \BibitemOpen
  \bibfield  {author} {\bibinfo {author} {\bibfnamefont {Georgios}\
  \bibnamefont {Kofinas}}, \bibinfo {author} {\bibfnamefont {Genly}\
  \bibnamefont {Leon}}, \ and\ \bibinfo {author} {\bibfnamefont {Emmanuel~N.}\
  \bibnamefont {Saridakis}},\ }\bibfield  {title} {\enquote {\bibinfo {title}
  {{Dynamical behavior in $f(T,T_G)$ cosmology}},}\ }\href {\doibase
  10.1088/0264-9381/31/17/175011} {\bibfield  {journal} {\bibinfo  {journal}
  {Class. Quant. Grav.}\ }\textbf {\bibinfo {volume} {31}},\ \bibinfo {pages}
  {175011} (\bibinfo {year} {2014})},\ \Eprint {http://arxiv.org/abs/1404.7100}
  {arXiv:1404.7100 [gr-qc]} \BibitemShut {NoStop}%
\bibitem [{\citenamefont {Kofinas}\ and\ \citenamefont
  {Saridakis}(2014{\natexlab{b}})}]{Kofinas:2014daa}%
  \BibitemOpen
  \bibfield  {author} {\bibinfo {author} {\bibfnamefont {Georgios}\
  \bibnamefont {Kofinas}}\ and\ \bibinfo {author} {\bibfnamefont {Emmanuel~N.}\
  \bibnamefont {Saridakis}},\ }\bibfield  {title} {\enquote {\bibinfo {title}
  {{Cosmological applications of $F(T,T_G)$ gravity}},}\ }\href {\doibase
  10.1103/PhysRevD.90.084045} {\bibfield  {journal} {\bibinfo  {journal} {Phys.
  Rev. D}\ }\textbf {\bibinfo {volume} {90}},\ \bibinfo {pages} {084045}
  (\bibinfo {year} {2014}{\natexlab{b}})},\ \Eprint
  {http://arxiv.org/abs/1408.0107} {arXiv:1408.0107 [gr-qc]} \BibitemShut
  {NoStop}%
\bibitem [{\citenamefont {Bahamonde}\ \emph {et~al.}(2015)\citenamefont
  {Bahamonde}, \citenamefont {B\"ohmer},\ and\ \citenamefont
  {Wright}}]{Bahamonde:2015zma}%
  \BibitemOpen
  \bibfield  {author} {\bibinfo {author} {\bibfnamefont {Sebastian}\
  \bibnamefont {Bahamonde}}, \bibinfo {author} {\bibfnamefont {Christian~G.}\
  \bibnamefont {B\"ohmer}}, \ and\ \bibinfo {author} {\bibfnamefont {Matthew}\
  \bibnamefont {Wright}},\ }\bibfield  {title} {\enquote {\bibinfo {title}
  {{Modified teleparallel theories of gravity}},}\ }\href {\doibase
  10.1103/PhysRevD.92.104042} {\bibfield  {journal} {\bibinfo  {journal} {Phys.
  Rev. D}\ }\textbf {\bibinfo {volume} {92}},\ \bibinfo {pages} {104042}
  (\bibinfo {year} {2015})},\ \Eprint {http://arxiv.org/abs/1508.05120}
  {arXiv:1508.05120 [gr-qc]} \BibitemShut {NoStop}%
\bibitem [{\citenamefont {Bahamonde}\ and\ \citenamefont
  {Capozziello}(2017)}]{Bahamonde:2016grb}%
  \BibitemOpen
  \bibfield  {author} {\bibinfo {author} {\bibfnamefont {Sebastian}\
  \bibnamefont {Bahamonde}}\ and\ \bibinfo {author} {\bibfnamefont {Salvatore}\
  \bibnamefont {Capozziello}},\ }\bibfield  {title} {\enquote {\bibinfo {title}
  {{Noether Symmetry Approach in $f(T,B)$ teleparallel cosmology}},}\ }\href
  {\doibase 10.1140/epjc/s10052-017-4677-0} {\bibfield  {journal} {\bibinfo
  {journal} {Eur. Phys. J. C}\ }\textbf {\bibinfo {volume} {77}},\ \bibinfo
  {pages} {107} (\bibinfo {year} {2017})},\ \Eprint
  {http://arxiv.org/abs/1612.01299} {arXiv:1612.01299 [gr-qc]} \BibitemShut
  {NoStop}%
\bibitem [{\citenamefont {Geng}\ \emph {et~al.}(2011)\citenamefont {Geng},
  \citenamefont {Lee}, \citenamefont {Saridakis},\ and\ \citenamefont
  {Wu}}]{Geng:2011aj}%
  \BibitemOpen
  \bibfield  {author} {\bibinfo {author} {\bibfnamefont {Chao-Qiang}\
  \bibnamefont {Geng}}, \bibinfo {author} {\bibfnamefont {Chung-Chi}\
  \bibnamefont {Lee}}, \bibinfo {author} {\bibfnamefont {Emmanuel~N.}\
  \bibnamefont {Saridakis}}, \ and\ \bibinfo {author} {\bibfnamefont {Yi-Peng}\
  \bibnamefont {Wu}},\ }\bibfield  {title} {\enquote {\bibinfo {title}
  {{\textquotedblleft{}Teleparallel\textquotedblright{} dark energy}},}\ }\href
  {\doibase 10.1016/j.physletb.2011.09.082} {\bibfield  {journal} {\bibinfo
  {journal} {Phys. Lett. B}\ }\textbf {\bibinfo {volume} {704}},\ \bibinfo
  {pages} {384--387} (\bibinfo {year} {2011})},\ \Eprint
  {http://arxiv.org/abs/1109.1092} {arXiv:1109.1092 [hep-th]} \BibitemShut
  {NoStop}%
\bibitem [{\citenamefont {Beltr\'an~Jim\'enez}\ \emph
  {et~al.}(2018)\citenamefont {Beltr\'an~Jim\'enez}, \citenamefont
  {Heisenberg},\ and\ \citenamefont {Koivisto}}]{BeltranJimenez:2017tkd}%
  \BibitemOpen
  \bibfield  {author} {\bibinfo {author} {\bibfnamefont {Jose}\ \bibnamefont
  {Beltr\'an~Jim\'enez}}, \bibinfo {author} {\bibfnamefont {Lavinia}\
  \bibnamefont {Heisenberg}}, \ and\ \bibinfo {author} {\bibfnamefont {Tomi}\
  \bibnamefont {Koivisto}},\ }\bibfield  {title} {\enquote {\bibinfo {title}
  {{Coincident General Relativity}},}\ }\href {\doibase
  10.1103/PhysRevD.98.044048} {\bibfield  {journal} {\bibinfo  {journal} {Phys.
  Rev. D}\ }\textbf {\bibinfo {volume} {98}},\ \bibinfo {pages} {044048}
  (\bibinfo {year} {2018})},\ \Eprint {http://arxiv.org/abs/1710.03116}
  {arXiv:1710.03116 [gr-qc]} \BibitemShut {NoStop}%
\bibitem [{\citenamefont {Beltr\'an~Jim\'enez}\ \emph
  {et~al.}(2020)\citenamefont {Beltr\'an~Jim\'enez}, \citenamefont
  {Heisenberg}, \citenamefont {Koivisto},\ and\ \citenamefont
  {Pekar}}]{Jimenez:2019ovq}%
  \BibitemOpen
  \bibfield  {author} {\bibinfo {author} {\bibfnamefont {Jose}\ \bibnamefont
  {Beltr\'an~Jim\'enez}}, \bibinfo {author} {\bibfnamefont {Lavinia}\
  \bibnamefont {Heisenberg}}, \bibinfo {author} {\bibfnamefont
  {Tomi~Sebastian}\ \bibnamefont {Koivisto}}, \ and\ \bibinfo {author}
  {\bibfnamefont {Simon}\ \bibnamefont {Pekar}},\ }\bibfield  {title} {\enquote
  {\bibinfo {title} {{Cosmology in $f(Q)$ geometry}},}\ }\href {\doibase
  10.1103/PhysRevD.101.103507} {\bibfield  {journal} {\bibinfo  {journal}
  {Phys. Rev. D}\ }\textbf {\bibinfo {volume} {101}},\ \bibinfo {pages}
  {103507} (\bibinfo {year} {2020})},\ \Eprint
  {http://arxiv.org/abs/1906.10027} {arXiv:1906.10027 [gr-qc]} \BibitemShut
  {NoStop}%
\bibitem [{\citenamefont {Dialektopoulos}\ \emph {et~al.}(2019)\citenamefont
  {Dialektopoulos}, \citenamefont {Koivisto},\ and\ \citenamefont
  {Capozziello}}]{Dialektopoulos:2019mtr}%
  \BibitemOpen
  \bibfield  {author} {\bibinfo {author} {\bibfnamefont {Konstantinos~F.}\
  \bibnamefont {Dialektopoulos}}, \bibinfo {author} {\bibfnamefont {Tomi~S.}\
  \bibnamefont {Koivisto}}, \ and\ \bibinfo {author} {\bibfnamefont
  {Salvatore}\ \bibnamefont {Capozziello}},\ }\bibfield  {title} {\enquote
  {\bibinfo {title} {{Noether symmetries in Symmetric Teleparallel
  Cosmology}},}\ }\href {\doibase 10.1140/epjc/s10052-019-7106-8} {\bibfield
  {journal} {\bibinfo  {journal} {Eur. Phys. J. C}\ }\textbf {\bibinfo {volume}
  {79}},\ \bibinfo {pages} {606} (\bibinfo {year} {2019})},\ \Eprint
  {http://arxiv.org/abs/1905.09019} {arXiv:1905.09019 [gr-qc]} \BibitemShut
  {NoStop}%
\bibitem [{\citenamefont {Bajardi}\ \emph {et~al.}(2020)\citenamefont
  {Bajardi}, \citenamefont {Vernieri},\ and\ \citenamefont
  {Capozziello}}]{Bajardi:2020fxh}%
  \BibitemOpen
  \bibfield  {author} {\bibinfo {author} {\bibfnamefont {Francesco}\
  \bibnamefont {Bajardi}}, \bibinfo {author} {\bibfnamefont {Daniele}\
  \bibnamefont {Vernieri}}, \ and\ \bibinfo {author} {\bibfnamefont
  {Salvatore}\ \bibnamefont {Capozziello}},\ }\bibfield  {title} {\enquote
  {\bibinfo {title} {{Bouncing Cosmology in f(Q) Symmetric Teleparallel
  Gravity}},}\ }\href {\doibase 10.1140/epjp/s13360-020-00918-3} {\bibfield
  {journal} {\bibinfo  {journal} {Eur. Phys. J. Plus}\ }\textbf {\bibinfo
  {volume} {135}},\ \bibinfo {pages} {912} (\bibinfo {year} {2020})},\ \Eprint
  {http://arxiv.org/abs/2011.01248} {arXiv:2011.01248 [gr-qc]} \BibitemShut
  {NoStop}%
\bibitem [{\citenamefont {Flathmann}\ and\ \citenamefont
  {Hohmann}(2021)}]{Flathmann:2020zyj}%
  \BibitemOpen
  \bibfield  {author} {\bibinfo {author} {\bibfnamefont {Kai}\ \bibnamefont
  {Flathmann}}\ and\ \bibinfo {author} {\bibfnamefont {Manuel}\ \bibnamefont
  {Hohmann}},\ }\bibfield  {title} {\enquote {\bibinfo {title} {{Post-Newtonian
  limit of generalized symmetric teleparallel gravity}},}\ }\href {\doibase
  10.1103/PhysRevD.103.044030} {\bibfield  {journal} {\bibinfo  {journal}
  {Phys. Rev. D}\ }\textbf {\bibinfo {volume} {103}},\ \bibinfo {pages}
  {044030} (\bibinfo {year} {2021})},\ \Eprint
  {http://arxiv.org/abs/2012.12875} {arXiv:2012.12875 [gr-qc]} \BibitemShut
  {NoStop}%
\bibitem [{\citenamefont {Mandal}\ \emph
  {et~al.}(2020{\natexlab{a}})\citenamefont {Mandal}, \citenamefont {Wang},\
  and\ \citenamefont {Sahoo}}]{Mandal:2020buf}%
  \BibitemOpen
  \bibfield  {author} {\bibinfo {author} {\bibfnamefont {Sanjay}\ \bibnamefont
  {Mandal}}, \bibinfo {author} {\bibfnamefont {Deng}\ \bibnamefont {Wang}}, \
  and\ \bibinfo {author} {\bibfnamefont {P.~K.}\ \bibnamefont {Sahoo}},\
  }\bibfield  {title} {\enquote {\bibinfo {title} {{Cosmography in $f(Q)$
  gravity}},}\ }\href {\doibase 10.1103/PhysRevD.102.124029} {\bibfield
  {journal} {\bibinfo  {journal} {Phys. Rev. D}\ }\textbf {\bibinfo {volume}
  {102}},\ \bibinfo {pages} {124029} (\bibinfo {year} {2020}{\natexlab{a}})},\
  \Eprint {http://arxiv.org/abs/2011.00420} {arXiv:2011.00420 [gr-qc]}
  \BibitemShut {NoStop}%
\bibitem [{\citenamefont {D'Ambrosio}\ \emph {et~al.}(2020)\citenamefont
  {D'Ambrosio}, \citenamefont {Garg},\ and\ \citenamefont
  {Heisenberg}}]{DAmbrosio:2020nev}%
  \BibitemOpen
  \bibfield  {author} {\bibinfo {author} {\bibfnamefont {Fabio}\ \bibnamefont
  {D'Ambrosio}}, \bibinfo {author} {\bibfnamefont {Mudit}\ \bibnamefont
  {Garg}}, \ and\ \bibinfo {author} {\bibfnamefont {Lavinia}\ \bibnamefont
  {Heisenberg}},\ }\bibfield  {title} {\enquote {\bibinfo {title} {{Non-linear
  extension of non-metricity scalar for MOND}},}\ }\href {\doibase
  10.1016/j.physletb.2020.135970} {\bibfield  {journal} {\bibinfo  {journal}
  {Phys. Lett. B}\ }\textbf {\bibinfo {volume} {811}},\ \bibinfo {pages}
  {135970} (\bibinfo {year} {2020})},\ \Eprint
  {http://arxiv.org/abs/2004.00888} {arXiv:2004.00888 [gr-qc]} \BibitemShut
  {NoStop}%
\bibitem [{\citenamefont {Mandal}\ \emph
  {et~al.}(2020{\natexlab{b}})\citenamefont {Mandal}, \citenamefont {Sahoo},\
  and\ \citenamefont {Santos}}]{Mandal:2020lyq}%
  \BibitemOpen
  \bibfield  {author} {\bibinfo {author} {\bibfnamefont {Sanjay}\ \bibnamefont
  {Mandal}}, \bibinfo {author} {\bibfnamefont {P.~K.}\ \bibnamefont {Sahoo}}, \
  and\ \bibinfo {author} {\bibfnamefont {J.~R.~L.}\ \bibnamefont {Santos}},\
  }\bibfield  {title} {\enquote {\bibinfo {title} {{Energy conditions in $f(Q)$
  gravity}},}\ }\href {\doibase 10.1103/PhysRevD.102.024057} {\bibfield
  {journal} {\bibinfo  {journal} {Phys. Rev. D}\ }\textbf {\bibinfo {volume}
  {102}},\ \bibinfo {pages} {024057} (\bibinfo {year} {2020}{\natexlab{b}})},\
  \Eprint {http://arxiv.org/abs/2008.01563} {arXiv:2008.01563 [gr-qc]}
  \BibitemShut {NoStop}%
\bibitem [{\citenamefont {Dimakis}\ \emph {et~al.}(2021)\citenamefont
  {Dimakis}, \citenamefont {Paliathanasis},\ and\ \citenamefont
  {Christodoulakis}}]{Dimakis:2021gby}%
  \BibitemOpen
  \bibfield  {author} {\bibinfo {author} {\bibfnamefont {N.}~\bibnamefont
  {Dimakis}}, \bibinfo {author} {\bibfnamefont {A.}~\bibnamefont
  {Paliathanasis}}, \ and\ \bibinfo {author} {\bibfnamefont {T.}~\bibnamefont
  {Christodoulakis}},\ }\bibfield  {title} {\enquote {\bibinfo {title}
  {{Quantum cosmology in f(Q) theory}},}\ }\href {\doibase
  10.1088/1361-6382/ac2b09} {\bibfield  {journal} {\bibinfo  {journal} {Class.
  Quant. Grav.}\ }\textbf {\bibinfo {volume} {38}},\ \bibinfo {pages} {225003}
  (\bibinfo {year} {2021})},\ \Eprint {http://arxiv.org/abs/2108.01970}
  {arXiv:2108.01970 [gr-qc]} \BibitemShut {NoStop}%
\bibitem [{\citenamefont {Nakayama}(2022)}]{Nakayama:2021rda}%
  \BibitemOpen
  \bibfield  {author} {\bibinfo {author} {\bibfnamefont {Yu}~\bibnamefont
  {Nakayama}},\ }\bibfield  {title} {\enquote {\bibinfo {title} {{Weyl
  transverse diffeomorphism invariant theory of symmetric teleparallel
  gravity}},}\ }\href {\doibase 10.1088/1361-6382/ac776b} {\bibfield  {journal}
  {\bibinfo  {journal} {Class. Quant. Grav.}\ }\textbf {\bibinfo {volume}
  {39}},\ \bibinfo {pages} {145006} (\bibinfo {year} {2022})},\ \Eprint
  {http://arxiv.org/abs/2108.10465} {arXiv:2108.10465 [gr-qc]} \BibitemShut
  {NoStop}%
\bibitem [{\citenamefont {Khyllep}\ \emph {et~al.}(2021)\citenamefont
  {Khyllep}, \citenamefont {Paliathanasis},\ and\ \citenamefont
  {Dutta}}]{Khyllep:2021pcu}%
  \BibitemOpen
  \bibfield  {author} {\bibinfo {author} {\bibfnamefont {Wompherdeiki}\
  \bibnamefont {Khyllep}}, \bibinfo {author} {\bibfnamefont {Andronikos}\
  \bibnamefont {Paliathanasis}}, \ and\ \bibinfo {author} {\bibfnamefont
  {Jibitesh}\ \bibnamefont {Dutta}},\ }\bibfield  {title} {\enquote {\bibinfo
  {title} {{Cosmological solutions and growth index of matter perturbations in
  $f(Q)$ gravity}},}\ }\href {\doibase 10.1103/PhysRevD.103.103521} {\bibfield
  {journal} {\bibinfo  {journal} {Phys. Rev. D}\ }\textbf {\bibinfo {volume}
  {103}},\ \bibinfo {pages} {103521} (\bibinfo {year} {2021})},\ \Eprint
  {http://arxiv.org/abs/2103.08372} {arXiv:2103.08372 [gr-qc]} \BibitemShut
  {NoStop}%
\bibitem [{\citenamefont {Hohmann}(2021)}]{Hohmann:2021ast}%
  \BibitemOpen
  \bibfield  {author} {\bibinfo {author} {\bibfnamefont {Manuel}\ \bibnamefont
  {Hohmann}},\ }\bibfield  {title} {\enquote {\bibinfo {title} {{General
  covariant symmetric teleparallel cosmology}},}\ }\href {\doibase
  10.1103/PhysRevD.104.124077} {\bibfield  {journal} {\bibinfo  {journal}
  {Phys. Rev. D}\ }\textbf {\bibinfo {volume} {104}},\ \bibinfo {pages}
  {124077} (\bibinfo {year} {2021})},\ \Eprint
  {http://arxiv.org/abs/2109.01525} {arXiv:2109.01525 [gr-qc]} \BibitemShut
  {NoStop}%
\bibitem [{\citenamefont {Wang}\ \emph {et~al.}(2022)\citenamefont {Wang},
  \citenamefont {Chen},\ and\ \citenamefont {Katsuragawa}}]{Wang:2021zaz}%
  \BibitemOpen
  \bibfield  {author} {\bibinfo {author} {\bibfnamefont {Wenyi}\ \bibnamefont
  {Wang}}, \bibinfo {author} {\bibfnamefont {Hua}\ \bibnamefont {Chen}}, \ and\
  \bibinfo {author} {\bibfnamefont {Taishi}\ \bibnamefont {Katsuragawa}},\
  }\bibfield  {title} {\enquote {\bibinfo {title} {{Static and spherically
  symmetric solutions in f(Q) gravity}},}\ }\href {\doibase
  10.1103/PhysRevD.105.024060} {\bibfield  {journal} {\bibinfo  {journal}
  {Phys. Rev. D}\ }\textbf {\bibinfo {volume} {105}},\ \bibinfo {pages}
  {024060} (\bibinfo {year} {2022})},\ \Eprint
  {http://arxiv.org/abs/2110.13565} {arXiv:2110.13565 [gr-qc]} \BibitemShut
  {NoStop}%
\bibitem [{\citenamefont {Quiros}(2022)}]{Quiros:2021eju}%
  \BibitemOpen
  \bibfield  {author} {\bibinfo {author} {\bibfnamefont {Israel}\ \bibnamefont
  {Quiros}},\ }\bibfield  {title} {\enquote {\bibinfo {title} {{Nonmetricity
  theories and aspects of gauge symmetry}},}\ }\href {\doibase
  10.1103/PhysRevD.105.104060} {\bibfield  {journal} {\bibinfo  {journal}
  {Phys. Rev. D}\ }\textbf {\bibinfo {volume} {105}},\ \bibinfo {pages}
  {104060} (\bibinfo {year} {2022})},\ \Eprint
  {http://arxiv.org/abs/2111.05490} {arXiv:2111.05490 [gr-qc]} \BibitemShut
  {NoStop}%
\bibitem [{\citenamefont {Ferreira}\ \emph {et~al.}(2022)\citenamefont
  {Ferreira}, \citenamefont {Barreiro}, \citenamefont {Mimoso},\ and\
  \citenamefont {Nunes}}]{Ferreira:2022jcd}%
  \BibitemOpen
  \bibfield  {author} {\bibinfo {author} {\bibfnamefont {Jos\'e}\ \bibnamefont
  {Ferreira}}, \bibinfo {author} {\bibfnamefont {Tiago}\ \bibnamefont
  {Barreiro}}, \bibinfo {author} {\bibfnamefont {Jos\'e}\ \bibnamefont
  {Mimoso}}, \ and\ \bibinfo {author} {\bibfnamefont {Nelson~J.}\ \bibnamefont
  {Nunes}},\ }\bibfield  {title} {\enquote {\bibinfo {title} {{Forecasting
  $F(Q)$ cosmology with $\Lambda$CDM background using standard sirens}},}\
  }\href@noop {} {\  (\bibinfo {year} {2022})},\ \Eprint
  {http://arxiv.org/abs/2203.13788} {arXiv:2203.13788 [astro-ph.CO]}
  \BibitemShut {NoStop}%
\bibitem [{\citenamefont {Solanki}\ \emph {et~al.}(2022)\citenamefont
  {Solanki}, \citenamefont {De},\ and\ \citenamefont
  {Sahoo}}]{Solanki:2022ccf}%
  \BibitemOpen
  \bibfield  {author} {\bibinfo {author} {\bibfnamefont {Raja}\ \bibnamefont
  {Solanki}}, \bibinfo {author} {\bibfnamefont {Avik}\ \bibnamefont {De}}, \
  and\ \bibinfo {author} {\bibfnamefont {P.~K.}\ \bibnamefont {Sahoo}},\
  }\bibfield  {title} {\enquote {\bibinfo {title} {{Complete dark energy
  scenario in $f(Q)$ gravity}},}\ }\href {\doibase 10.1016/j.dark.2022.100996}
  {\bibfield  {journal} {\bibinfo  {journal} {Phys. Dark Univ.}\ }\textbf
  {\bibinfo {volume} {36}},\ \bibinfo {pages} {100996} (\bibinfo {year}
  {2022})},\ \Eprint {http://arxiv.org/abs/2203.03370} {arXiv:2203.03370
  [gr-qc]} \BibitemShut {NoStop}%
\bibitem [{\citenamefont {De}\ \emph {et~al.}(2022)\citenamefont {De},
  \citenamefont {Mandal}, \citenamefont {Beh}, \citenamefont {Loo},\ and\
  \citenamefont {Sahoo}}]{De:2022shr}%
  \BibitemOpen
  \bibfield  {author} {\bibinfo {author} {\bibfnamefont {Avik}\ \bibnamefont
  {De}}, \bibinfo {author} {\bibfnamefont {Sanjay}\ \bibnamefont {Mandal}},
  \bibinfo {author} {\bibfnamefont {J.~T.}\ \bibnamefont {Beh}}, \bibinfo
  {author} {\bibfnamefont {Tee-How}\ \bibnamefont {Loo}}, \ and\ \bibinfo
  {author} {\bibfnamefont {P.~K.}\ \bibnamefont {Sahoo}},\ }\bibfield  {title}
  {\enquote {\bibinfo {title} {{Isotropization of locally rotationally
  symmetric Bianchi-I universe in $f(Q)$-gravity}},}\ }\href {\doibase
  10.1140/epjc/s10052-022-10021-9} {\bibfield  {journal} {\bibinfo  {journal}
  {Eur. Phys. J. C}\ }\textbf {\bibinfo {volume} {82}},\ \bibinfo {pages} {72}
  (\bibinfo {year} {2022})},\ \Eprint {http://arxiv.org/abs/2201.05036}
  {arXiv:2201.05036 [gr-qc]} \BibitemShut {NoStop}%
\bibitem [{\citenamefont {Solanki}\ \emph {et~al.}(2021)\citenamefont
  {Solanki}, \citenamefont {Pacif}, \citenamefont {Parida},\ and\ \citenamefont
  {Sahoo}}]{Solanki:2021qni}%
  \BibitemOpen
  \bibfield  {author} {\bibinfo {author} {\bibfnamefont {Raja}\ \bibnamefont
  {Solanki}}, \bibinfo {author} {\bibfnamefont {S.~K.~J.}\ \bibnamefont
  {Pacif}}, \bibinfo {author} {\bibfnamefont {Abhishek}\ \bibnamefont
  {Parida}}, \ and\ \bibinfo {author} {\bibfnamefont {P.~K.}\ \bibnamefont
  {Sahoo}},\ }\bibfield  {title} {\enquote {\bibinfo {title} {{Cosmic
  acceleration with bulk viscosity in modified f(Q) gravity}},}\ }\href
  {\doibase 10.1016/j.dark.2021.100820} {\bibfield  {journal} {\bibinfo
  {journal} {Phys. Dark Univ.}\ }\textbf {\bibinfo {volume} {32}},\ \bibinfo
  {pages} {100820} (\bibinfo {year} {2021})},\ \Eprint
  {http://arxiv.org/abs/2105.00876} {arXiv:2105.00876 [gr-qc]} \BibitemShut
  {NoStop}%
\bibitem [{\citenamefont {Capozziello}\ and\ \citenamefont
  {D'Agostino}(2022)}]{Capozziello:2022wgl}%
  \BibitemOpen
  \bibfield  {author} {\bibinfo {author} {\bibfnamefont {Salvatore}\
  \bibnamefont {Capozziello}}\ and\ \bibinfo {author} {\bibfnamefont {Rocco}\
  \bibnamefont {D'Agostino}},\ }\bibfield  {title} {\enquote {\bibinfo {title}
  {{Model-independent reconstruction of f(Q) non-metric gravity}},}\ }\href
  {\doibase 10.1016/j.physletb.2022.137229} {\bibfield  {journal} {\bibinfo
  {journal} {Phys. Lett. B}\ }\textbf {\bibinfo {volume} {832}},\ \bibinfo
  {pages} {137229} (\bibinfo {year} {2022})},\ \Eprint
  {http://arxiv.org/abs/2204.01015} {arXiv:2204.01015 [gr-qc]} \BibitemShut
  {NoStop}%
\bibitem [{\citenamefont {Narawade}\ \emph {et~al.}(2022)\citenamefont
  {Narawade}, \citenamefont {Pati}, \citenamefont {Mishra},\ and\ \citenamefont
  {Tripathy}}]{Narawade:2022jeg}%
  \BibitemOpen
  \bibfield  {author} {\bibinfo {author} {\bibfnamefont {S.~A.}\ \bibnamefont
  {Narawade}}, \bibinfo {author} {\bibfnamefont {Laxmipriya}\ \bibnamefont
  {Pati}}, \bibinfo {author} {\bibfnamefont {B.}~\bibnamefont {Mishra}}, \ and\
  \bibinfo {author} {\bibfnamefont {S.~K.}\ \bibnamefont {Tripathy}},\
  }\bibfield  {title} {\enquote {\bibinfo {title} {{Dynamical system analysis
  for accelerating models in non-metricity f(Q) gravity}},}\ }\href {\doibase
  10.1016/j.dark.2022.101020} {\bibfield  {journal} {\bibinfo  {journal} {Phys.
  Dark Univ.}\ }\textbf {\bibinfo {volume} {36}},\ \bibinfo {pages} {101020}
  (\bibinfo {year} {2022})},\ \Eprint {http://arxiv.org/abs/2203.14121}
  {arXiv:2203.14121 [gr-qc]} \BibitemShut {NoStop}%
\bibitem [{\citenamefont {Dimakis}\ \emph {et~al.}(2022)\citenamefont
  {Dimakis}, \citenamefont {Paliathanasis}, \citenamefont {Roumeliotis},\ and\
  \citenamefont {Christodoulakis}}]{Dimakis:2022rkd}%
  \BibitemOpen
  \bibfield  {author} {\bibinfo {author} {\bibfnamefont {N.}~\bibnamefont
  {Dimakis}}, \bibinfo {author} {\bibfnamefont {A.}~\bibnamefont
  {Paliathanasis}}, \bibinfo {author} {\bibfnamefont {M.}~\bibnamefont
  {Roumeliotis}}, \ and\ \bibinfo {author} {\bibfnamefont {T.}~\bibnamefont
  {Christodoulakis}},\ }\bibfield  {title} {\enquote {\bibinfo {title} {{FLRW
  solutions in $f(Q)$ theory: the effect of using different connections}},}\
  }\href@noop {} {\  (\bibinfo {year} {2022})},\ \Eprint
  {http://arxiv.org/abs/2205.04680} {arXiv:2205.04680 [gr-qc]} \BibitemShut
  {NoStop}%
\bibitem [{\citenamefont {Albuquerque}\ and\ \citenamefont
  {Frusciante}(2022)}]{Albuquerque:2022eac}%
  \BibitemOpen
  \bibfield  {author} {\bibinfo {author} {\bibfnamefont {In\^es~S.}\
  \bibnamefont {Albuquerque}}\ and\ \bibinfo {author} {\bibfnamefont {Noemi}\
  \bibnamefont {Frusciante}},\ }\bibfield  {title} {\enquote {\bibinfo {title}
  {{A designer approach to f(Q) gravity and cosmological implications}},}\
  }\href {\doibase 10.1016/j.dark.2022.100980} {\bibfield  {journal} {\bibinfo
  {journal} {Phys. Dark Univ.}\ }\textbf {\bibinfo {volume} {35}},\ \bibinfo
  {pages} {100980} (\bibinfo {year} {2022})},\ \Eprint
  {http://arxiv.org/abs/2202.04637} {arXiv:2202.04637 [astro-ph.CO]}
  \BibitemShut {NoStop}%
\bibitem [{\citenamefont {Arora}\ and\ \citenamefont
  {Sahoo}(2022)}]{Arora:2022mlo}%
  \BibitemOpen
  \bibfield  {author} {\bibinfo {author} {\bibfnamefont {Simran}\ \bibnamefont
  {Arora}}\ and\ \bibinfo {author} {\bibfnamefont {P.~K.}\ \bibnamefont
  {Sahoo}},\ }\bibfield  {title} {\enquote {\bibinfo {title} {{Crossing phantom
  divide in $f(Q)$ gravity}},}\ }\href {\doibase 10.1002/andp.202200233} {\
  (\bibinfo {year} {2022}),\ 10.1002/andp.202200233},\ \Eprint
  {http://arxiv.org/abs/2206.05110} {arXiv:2206.05110 [gr-qc]} \BibitemShut
  {NoStop}%
\bibitem [{\citenamefont {Pati}\ \emph {et~al.}(2022)\citenamefont {Pati},
  \citenamefont {Narawade}, \citenamefont {Tripathy},\ and\ \citenamefont
  {Mishra}}]{Pati:2022dwl}%
  \BibitemOpen
  \bibfield  {author} {\bibinfo {author} {\bibfnamefont {Laxmipriya}\
  \bibnamefont {Pati}}, \bibinfo {author} {\bibfnamefont {S.~A.}\ \bibnamefont
  {Narawade}}, \bibinfo {author} {\bibfnamefont {S.~K.}\ \bibnamefont
  {Tripathy}}, \ and\ \bibinfo {author} {\bibfnamefont {B.}~\bibnamefont
  {Mishra}},\ }\bibfield  {title} {\enquote {\bibinfo {title} {{Scalar
  perturbations in a class of extended symmetric teleparallel gravity
  models}},}\ }\href@noop {} {\  (\bibinfo {year} {2022})},\ \Eprint
  {http://arxiv.org/abs/2206.11928} {arXiv:2206.11928 [gr-qc]} \BibitemShut
  {NoStop}%
\bibitem [{\citenamefont {Soudi}\ \emph {et~al.}(2019)\citenamefont {Soudi},
  \citenamefont {Farrugia}, \citenamefont {Gakis}, \citenamefont {Levi~Said},\
  and\ \citenamefont {Saridakis}}]{Soudi:2018dhv}%
  \BibitemOpen
  \bibfield  {author} {\bibinfo {author} {\bibfnamefont {Ismail}\ \bibnamefont
  {Soudi}}, \bibinfo {author} {\bibfnamefont {Gabriel}\ \bibnamefont
  {Farrugia}}, \bibinfo {author} {\bibfnamefont {Viktor}\ \bibnamefont
  {Gakis}}, \bibinfo {author} {\bibfnamefont {Jackson}\ \bibnamefont
  {Levi~Said}}, \ and\ \bibinfo {author} {\bibfnamefont {Emmanuel~N.}\
  \bibnamefont {Saridakis}},\ }\bibfield  {title} {\enquote {\bibinfo {title}
  {{Polarization of gravitational waves in symmetric teleparallel theories of
  gravity and their modifications}},}\ }\href {\doibase
  10.1103/PhysRevD.100.044008} {\bibfield  {journal} {\bibinfo  {journal}
  {Phys. Rev. D}\ }\textbf {\bibinfo {volume} {100}},\ \bibinfo {pages}
  {044008} (\bibinfo {year} {2019})},\ \Eprint
  {http://arxiv.org/abs/1810.08220} {arXiv:1810.08220 [gr-qc]} \BibitemShut
  {NoStop}%
\bibitem [{\citenamefont {Lazkoz}\ \emph {et~al.}(2019)\citenamefont {Lazkoz},
  \citenamefont {Lobo}, \citenamefont {Ortiz-Ba\~nos},\ and\ \citenamefont
  {Salzano}}]{Lazkoz:2019sjl}%
  \BibitemOpen
  \bibfield  {author} {\bibinfo {author} {\bibfnamefont {Ruth}\ \bibnamefont
  {Lazkoz}}, \bibinfo {author} {\bibfnamefont {Francisco S.~N.}\ \bibnamefont
  {Lobo}}, \bibinfo {author} {\bibfnamefont {Mar\'\i{}a}\ \bibnamefont
  {Ortiz-Ba\~nos}}, \ and\ \bibinfo {author} {\bibfnamefont {Vincenzo}\
  \bibnamefont {Salzano}},\ }\bibfield  {title} {\enquote {\bibinfo {title}
  {{Observational constraints of $f(Q)$ gravity}},}\ }\href {\doibase
  10.1103/PhysRevD.100.104027} {\bibfield  {journal} {\bibinfo  {journal}
  {Phys. Rev. D}\ }\textbf {\bibinfo {volume} {100}},\ \bibinfo {pages}
  {104027} (\bibinfo {year} {2019})},\ \Eprint
  {http://arxiv.org/abs/1907.13219} {arXiv:1907.13219 [gr-qc]} \BibitemShut
  {NoStop}%
\bibitem [{\citenamefont {Barros}\ \emph {et~al.}(2020)\citenamefont {Barros},
  \citenamefont {Barreiro}, \citenamefont {Koivisto},\ and\ \citenamefont
  {Nunes}}]{Barros:2020bgg}%
  \BibitemOpen
  \bibfield  {author} {\bibinfo {author} {\bibfnamefont {Bruno~J.}\
  \bibnamefont {Barros}}, \bibinfo {author} {\bibfnamefont {Tiago}\
  \bibnamefont {Barreiro}}, \bibinfo {author} {\bibfnamefont {Tomi}\
  \bibnamefont {Koivisto}}, \ and\ \bibinfo {author} {\bibfnamefont
  {Nelson~J.}\ \bibnamefont {Nunes}},\ }\bibfield  {title} {\enquote {\bibinfo
  {title} {{Testing $F(Q)$ gravity with redshift space distortions}},}\ }\href
  {\doibase 10.1016/j.dark.2020.100616} {\bibfield  {journal} {\bibinfo
  {journal} {Phys. Dark Univ.}\ }\textbf {\bibinfo {volume} {30}},\ \bibinfo
  {pages} {100616} (\bibinfo {year} {2020})},\ \Eprint
  {http://arxiv.org/abs/2004.07867} {arXiv:2004.07867 [gr-qc]} \BibitemShut
  {NoStop}%
\bibitem [{\citenamefont {Ayuso}\ \emph {et~al.}(2021)\citenamefont {Ayuso},
  \citenamefont {Lazkoz},\ and\ \citenamefont {Salzano}}]{Ayuso:2020dcu}%
  \BibitemOpen
  \bibfield  {author} {\bibinfo {author} {\bibfnamefont {Ismael}\ \bibnamefont
  {Ayuso}}, \bibinfo {author} {\bibfnamefont {Ruth}\ \bibnamefont {Lazkoz}}, \
  and\ \bibinfo {author} {\bibfnamefont {Vincenzo}\ \bibnamefont {Salzano}},\
  }\bibfield  {title} {\enquote {\bibinfo {title} {{Observational constraints
  on cosmological solutions of $f(Q)$ theories}},}\ }\href {\doibase
  10.1103/PhysRevD.103.063505} {\bibfield  {journal} {\bibinfo  {journal}
  {Phys. Rev. D}\ }\textbf {\bibinfo {volume} {103}},\ \bibinfo {pages}
  {063505} (\bibinfo {year} {2021})},\ \Eprint
  {http://arxiv.org/abs/2012.00046} {arXiv:2012.00046 [astro-ph.CO]}
  \BibitemShut {NoStop}%
\bibitem [{\citenamefont {Anagnostopoulos}\ \emph {et~al.}(2021)\citenamefont
  {Anagnostopoulos}, \citenamefont {Basilakos},\ and\ \citenamefont
  {Saridakis}}]{Anagnostopoulos:2021ydo}%
  \BibitemOpen
  \bibfield  {author} {\bibinfo {author} {\bibfnamefont {Fotios~K.}\
  \bibnamefont {Anagnostopoulos}}, \bibinfo {author} {\bibfnamefont {Spyros}\
  \bibnamefont {Basilakos}}, \ and\ \bibinfo {author} {\bibfnamefont
  {Emmanuel~N.}\ \bibnamefont {Saridakis}},\ }\bibfield  {title} {\enquote
  {\bibinfo {title} {{First evidence that non-metricity f(Q) gravity could
  challenge \ensuremath{\Lambda}CDM}},}\ }\href {\doibase
  10.1016/j.physletb.2021.136634} {\bibfield  {journal} {\bibinfo  {journal}
  {Phys. Lett. B}\ }\textbf {\bibinfo {volume} {822}},\ \bibinfo {pages}
  {136634} (\bibinfo {year} {2021})},\ \Eprint
  {http://arxiv.org/abs/2104.15123} {arXiv:2104.15123 [gr-qc]} \BibitemShut
  {NoStop}%
\bibitem [{\citenamefont {Mandal}\ and\ \citenamefont
  {Sahoo}(2021)}]{Mandal:2021bpd}%
  \BibitemOpen
  \bibfield  {author} {\bibinfo {author} {\bibfnamefont {Sanjay}\ \bibnamefont
  {Mandal}}\ and\ \bibinfo {author} {\bibfnamefont {P.~K.}\ \bibnamefont
  {Sahoo}},\ }\bibfield  {title} {\enquote {\bibinfo {title} {{Constraint on
  the equation of state parameter ($\omega$) in non-minimally coupled $f(Q)$
  gravity}},}\ }\href {\doibase 10.1016/j.physletb.2021.136786} {\bibfield
  {journal} {\bibinfo  {journal} {Phys. Lett. B}\ }\textbf {\bibinfo {volume}
  {823}},\ \bibinfo {pages} {136786} (\bibinfo {year} {2021})},\ \Eprint
  {http://arxiv.org/abs/2111.10511} {arXiv:2111.10511 [gr-qc]} \BibitemShut
  {NoStop}%
\bibitem [{\citenamefont {Atayde}\ and\ \citenamefont
  {Frusciante}(2021)}]{Atayde:2021pgb}%
  \BibitemOpen
  \bibfield  {author} {\bibinfo {author} {\bibfnamefont {Lu\'\i{}s}\
  \bibnamefont {Atayde}}\ and\ \bibinfo {author} {\bibfnamefont {Noemi}\
  \bibnamefont {Frusciante}},\ }\bibfield  {title} {\enquote {\bibinfo {title}
  {{Can $f(Q)$ gravity challenge $\Lambda$CDM?}}}\ }\href {\doibase
  10.1103/PhysRevD.104.064052} {\bibfield  {journal} {\bibinfo  {journal}
  {Phys. Rev. D}\ }\textbf {\bibinfo {volume} {104}},\ \bibinfo {pages}
  {064052} (\bibinfo {year} {2021})},\ \Eprint
  {http://arxiv.org/abs/2108.10832} {arXiv:2108.10832 [astro-ph.CO]}
  \BibitemShut {NoStop}%
\bibitem [{\citenamefont {Frusciante}(2021)}]{Frusciante:2021sio}%
  \BibitemOpen
  \bibfield  {author} {\bibinfo {author} {\bibfnamefont {Noemi}\ \bibnamefont
  {Frusciante}},\ }\bibfield  {title} {\enquote {\bibinfo {title} {{Signatures
  of $f(Q)$-gravity in cosmology}},}\ }\href {\doibase
  10.1103/PhysRevD.103.044021} {\bibfield  {journal} {\bibinfo  {journal}
  {Phys. Rev. D}\ }\textbf {\bibinfo {volume} {103}},\ \bibinfo {pages}
  {044021} (\bibinfo {year} {2021})},\ \Eprint
  {http://arxiv.org/abs/2101.09242} {arXiv:2101.09242 [astro-ph.CO]}
  \BibitemShut {NoStop}%
\bibitem [{\citenamefont {Anagnostopoulos}\ \emph {et~al.}(2022)\citenamefont
  {Anagnostopoulos}, \citenamefont {Gakis}, \citenamefont {Saridakis},\ and\
  \citenamefont {Basilakos}}]{Anagnostopoulos:2022gej}%
  \BibitemOpen
  \bibfield  {author} {\bibinfo {author} {\bibfnamefont {Fotios~K.}\
  \bibnamefont {Anagnostopoulos}}, \bibinfo {author} {\bibfnamefont {Viktor}\
  \bibnamefont {Gakis}}, \bibinfo {author} {\bibfnamefont {Emmanuel~N.}\
  \bibnamefont {Saridakis}}, \ and\ \bibinfo {author} {\bibfnamefont {Spyros}\
  \bibnamefont {Basilakos}},\ }\bibfield  {title} {\enquote {\bibinfo {title}
  {{New models and Big Bang Nucleosynthesis constraints in $f(Q)$ gravity}},}\
  }\href@noop {} {\  (\bibinfo {year} {2022})},\ \Eprint
  {http://arxiv.org/abs/2205.11445} {arXiv:2205.11445 [gr-qc]} \BibitemShut
  {NoStop}%
\bibitem [{\citenamefont {Wainwright}\ and\ \citenamefont
  {Ellis}(1997)}]{wainwrightellis1997}%
  \BibitemOpen
  \bibfield  {author} {\bibinfo {author} {\bibfnamefont {John}\ \bibnamefont
  {Wainwright}}\ and\ \bibinfo {author} {\bibfnamefont {George F.~R.}\
  \bibnamefont {Ellis}},\ }\href {\doibase 10.1017/CBO9780511524660} {\emph
  {\bibinfo {title} {Dynamical Systems in Cosmology}}}\ (\bibinfo  {publisher}
  {Cambridge University Press (Cambridge)},\ \bibinfo {year}
  {1997})\BibitemShut {NoStop}%
\bibitem [{\citenamefont {Coley}(2003)}]{Coley:2003mj}%
  \BibitemOpen
  \bibfield  {author} {\bibinfo {author} {\bibfnamefont {A.~A.}\ \bibnamefont
  {Coley}},\ }\href {\doibase 10.1007/978-94-017-0327-7} {\emph {\bibinfo
  {title} {{Dynamical systems and cosmology}}}}\ (\bibinfo  {publisher}
  {Kluwer},\ \bibinfo {address} {Dordrecht, Netherlands},\ \bibinfo {year}
  {2003})\BibitemShut {NoStop}%
\bibitem [{\citenamefont {Bahamonde}\ \emph {et~al.}(2018)\citenamefont
  {Bahamonde}, \citenamefont {B\"ohmer}, \citenamefont {Carloni}, \citenamefont
  {Copeland}, \citenamefont {Fang},\ and\ \citenamefont
  {Tamanini}}]{Bahamonde:2017ize}%
  \BibitemOpen
  \bibfield  {author} {\bibinfo {author} {\bibfnamefont {Sebastian}\
  \bibnamefont {Bahamonde}}, \bibinfo {author} {\bibfnamefont {Christian~G.}\
  \bibnamefont {B\"ohmer}}, \bibinfo {author} {\bibfnamefont {Sante}\
  \bibnamefont {Carloni}}, \bibinfo {author} {\bibfnamefont {Edmund~J.}\
  \bibnamefont {Copeland}}, \bibinfo {author} {\bibfnamefont {Wei}\
  \bibnamefont {Fang}}, \ and\ \bibinfo {author} {\bibfnamefont {Nicola}\
  \bibnamefont {Tamanini}},\ }\bibfield  {title} {\enquote {\bibinfo {title}
  {{Dynamical systems applied to cosmology: dark energy and modified
  gravity}},}\ }\href {\doibase 10.1016/j.physrep.2018.09.001} {\bibfield
  {journal} {\bibinfo  {journal} {Phys. Rept.}\ }\textbf {\bibinfo {volume}
  {775-777}},\ \bibinfo {pages} {1--122} (\bibinfo {year} {2018})},\ \Eprint
  {http://arxiv.org/abs/1712.03107} {arXiv:1712.03107 [gr-qc]} \BibitemShut
  {NoStop}%
\bibitem [{\citenamefont {Copeland}\ \emph {et~al.}(1998)\citenamefont
  {Copeland}, \citenamefont {Liddle},\ and\ \citenamefont
  {Wands}}]{Copeland:1997et}%
  \BibitemOpen
  \bibfield  {author} {\bibinfo {author} {\bibfnamefont {Edmund~J.}\
  \bibnamefont {Copeland}}, \bibinfo {author} {\bibfnamefont {Andrew~R}\
  \bibnamefont {Liddle}}, \ and\ \bibinfo {author} {\bibfnamefont {David}\
  \bibnamefont {Wands}},\ }\bibfield  {title} {\enquote {\bibinfo {title}
  {{Exponential potentials and cosmological scaling solutions}},}\ }\href
  {\doibase 10.1103/PhysRevD.57.4686} {\bibfield  {journal} {\bibinfo
  {journal} {Phys. Rev. D}\ }\textbf {\bibinfo {volume} {57}},\ \bibinfo
  {pages} {4686--4690} (\bibinfo {year} {1998})},\ \Eprint
  {http://arxiv.org/abs/gr-qc/9711068} {arXiv:gr-qc/9711068} \BibitemShut
  {NoStop}%
\bibitem [{\citenamefont {Gong}\ \emph {et~al.}(2006)\citenamefont {Gong},
  \citenamefont {Wang},\ and\ \citenamefont {Zhang}}]{Gong:2006sp}%
  \BibitemOpen
  \bibfield  {author} {\bibinfo {author} {\bibfnamefont {Yungui}\ \bibnamefont
  {Gong}}, \bibinfo {author} {\bibfnamefont {Anzhong}\ \bibnamefont {Wang}}, \
  and\ \bibinfo {author} {\bibfnamefont {Yuan-Zhong}\ \bibnamefont {Zhang}},\
  }\bibfield  {title} {\enquote {\bibinfo {title} {{Exact scaling solutions and
  fixed points for general scalar field}},}\ }\href {\doibase
  10.1016/j.physletb.2006.03.057} {\bibfield  {journal} {\bibinfo  {journal}
  {Phys. Lett. B}\ }\textbf {\bibinfo {volume} {636}},\ \bibinfo {pages}
  {286--292} (\bibinfo {year} {2006})},\ \Eprint
  {http://arxiv.org/abs/gr-qc/0603050} {arXiv:gr-qc/0603050} \BibitemShut
  {NoStop}%
\bibitem [{\citenamefont {Setare}\ and\ \citenamefont
  {Saridakis}(2009)}]{Setare:2008sf}%
  \BibitemOpen
  \bibfield  {author} {\bibinfo {author} {\bibfnamefont {M.~R.}\ \bibnamefont
  {Setare}}\ and\ \bibinfo {author} {\bibfnamefont {E.~N.}\ \bibnamefont
  {Saridakis}},\ }\bibfield  {title} {\enquote {\bibinfo {title} {{Quintom dark
  energy models with nearly flat potentials}},}\ }\href {\doibase
  10.1103/PhysRevD.79.043005} {\bibfield  {journal} {\bibinfo  {journal} {Phys.
  Rev. D}\ }\textbf {\bibinfo {volume} {79}},\ \bibinfo {pages} {043005}
  (\bibinfo {year} {2009})},\ \Eprint {http://arxiv.org/abs/0810.4775}
  {arXiv:0810.4775 [astro-ph]} \BibitemShut {NoStop}%
\bibitem [{\citenamefont {Matos}\ \emph {et~al.}(2009)\citenamefont {Matos},
  \citenamefont {Luevano}, \citenamefont {Quiros}, \citenamefont
  {Urena-Lopez},\ and\ \citenamefont {Vazquez}}]{Matos:2009hf}%
  \BibitemOpen
  \bibfield  {author} {\bibinfo {author} {\bibfnamefont {Tonatiuh}\
  \bibnamefont {Matos}}, \bibinfo {author} {\bibfnamefont {Jose-Ruben}\
  \bibnamefont {Luevano}}, \bibinfo {author} {\bibfnamefont {Israel}\
  \bibnamefont {Quiros}}, \bibinfo {author} {\bibfnamefont {L.~Arturo}\
  \bibnamefont {Urena-Lopez}}, \ and\ \bibinfo {author} {\bibfnamefont
  {Jose~Alberto}\ \bibnamefont {Vazquez}},\ }\bibfield  {title} {\enquote
  {\bibinfo {title} {{Dynamics of Scalar Field Dark Matter With a Cosh-like
  Potential}},}\ }\href {\doibase 10.1103/PhysRevD.80.123521} {\bibfield
  {journal} {\bibinfo  {journal} {Phys. Rev. D}\ }\textbf {\bibinfo {volume}
  {80}},\ \bibinfo {pages} {123521} (\bibinfo {year} {2009})},\ \Eprint
  {http://arxiv.org/abs/0906.0396} {arXiv:0906.0396 [astro-ph.CO]} \BibitemShut
  {NoStop}%
\bibitem [{\citenamefont {Copeland}\ \emph {et~al.}(2009)\citenamefont
  {Copeland}, \citenamefont {Mizuno},\ and\ \citenamefont
  {Shaeri}}]{Copeland:2009be}%
  \BibitemOpen
  \bibfield  {author} {\bibinfo {author} {\bibfnamefont {Edmund~J.}\
  \bibnamefont {Copeland}}, \bibinfo {author} {\bibfnamefont {Shuntaro}\
  \bibnamefont {Mizuno}}, \ and\ \bibinfo {author} {\bibfnamefont {Maryam}\
  \bibnamefont {Shaeri}},\ }\bibfield  {title} {\enquote {\bibinfo {title}
  {{Dynamics of a scalar field in Robertson-Walker spacetimes}},}\ }\href
  {\doibase 10.1103/PhysRevD.79.103515} {\bibfield  {journal} {\bibinfo
  {journal} {Phys. Rev. D}\ }\textbf {\bibinfo {volume} {79}},\ \bibinfo
  {pages} {103515} (\bibinfo {year} {2009})},\ \Eprint
  {http://arxiv.org/abs/0904.0877} {arXiv:0904.0877 [astro-ph.CO]} \BibitemShut
  {NoStop}%
\bibitem [{\citenamefont {Leyva}\ \emph {et~al.}(2009)\citenamefont {Leyva},
  \citenamefont {Gonzalez}, \citenamefont {Gonzalez}, \citenamefont {Matos},\
  and\ \citenamefont {Quiros}}]{Leyva:2009zz}%
  \BibitemOpen
  \bibfield  {author} {\bibinfo {author} {\bibfnamefont {Yoelsy}\ \bibnamefont
  {Leyva}}, \bibinfo {author} {\bibfnamefont {Dania}\ \bibnamefont {Gonzalez}},
  \bibinfo {author} {\bibfnamefont {Tame}\ \bibnamefont {Gonzalez}}, \bibinfo
  {author} {\bibfnamefont {Tonatiuh}\ \bibnamefont {Matos}}, \ and\ \bibinfo
  {author} {\bibfnamefont {Israel}\ \bibnamefont {Quiros}},\ }\bibfield
  {title} {\enquote {\bibinfo {title} {{Dynamics of a self-interacting scalar
  field trapped in the braneworld for a wide variety of self-interaction
  potentials}},}\ }\href {\doibase 10.1103/PhysRevD.80.044026} {\bibfield
  {journal} {\bibinfo  {journal} {Phys. Rev. D}\ }\textbf {\bibinfo {volume}
  {80}},\ \bibinfo {pages} {044026} (\bibinfo {year} {2009})},\ \Eprint
  {http://arxiv.org/abs/0909.0281} {arXiv:0909.0281 [gr-qc]} \BibitemShut
  {NoStop}%
\bibitem [{\citenamefont {Leon}\ and\ \citenamefont
  {Saridakis}(2011)}]{Leon:2010pu}%
  \BibitemOpen
  \bibfield  {author} {\bibinfo {author} {\bibfnamefont {Genly}\ \bibnamefont
  {Leon}}\ and\ \bibinfo {author} {\bibfnamefont {Emmanuel~N.}\ \bibnamefont
  {Saridakis}},\ }\bibfield  {title} {\enquote {\bibinfo {title} {{Dynamics of
  the anisotropic Kantowsky-Sachs geometries in $R^n$ gravity}},}\ }\href
  {\doibase 10.1088/0264-9381/28/6/065008} {\bibfield  {journal} {\bibinfo
  {journal} {Class. Quant. Grav.}\ }\textbf {\bibinfo {volume} {28}},\ \bibinfo
  {pages} {065008} (\bibinfo {year} {2011})},\ \Eprint
  {http://arxiv.org/abs/1007.3956} {arXiv:1007.3956 [gr-qc]} \BibitemShut
  {NoStop}%
\bibitem [{\citenamefont {Urena-Lopez}(2012)}]{Urena-Lopez:2011gxx}%
  \BibitemOpen
  \bibfield  {author} {\bibinfo {author} {\bibfnamefont {L.~Arturo}\
  \bibnamefont {Urena-Lopez}},\ }\bibfield  {title} {\enquote {\bibinfo {title}
  {{Unified description of the dynamics of quintessential scalar fields}},}\
  }\href {\doibase 10.1088/1475-7516/2012/03/035} {\bibfield  {journal}
  {\bibinfo  {journal} {JCAP}\ }\textbf {\bibinfo {volume} {03}},\ \bibinfo
  {pages} {035} (\bibinfo {year} {2012})},\ \Eprint
  {http://arxiv.org/abs/1108.4712} {arXiv:1108.4712 [astro-ph.CO]} \BibitemShut
  {NoStop}%
\bibitem [{\citenamefont {Leon}\ \emph {et~al.}(2013)\citenamefont {Leon},
  \citenamefont {Saavedra},\ and\ \citenamefont {Saridakis}}]{Leon:2013qh}%
  \BibitemOpen
  \bibfield  {author} {\bibinfo {author} {\bibfnamefont {Genly}\ \bibnamefont
  {Leon}}, \bibinfo {author} {\bibfnamefont {Joel}\ \bibnamefont {Saavedra}}, \
  and\ \bibinfo {author} {\bibfnamefont {Emmanuel~N.}\ \bibnamefont
  {Saridakis}},\ }\bibfield  {title} {\enquote {\bibinfo {title} {{Cosmological
  behavior in extended nonlinear massive gravity}},}\ }\href {\doibase
  10.1088/0264-9381/30/13/135001} {\bibfield  {journal} {\bibinfo  {journal}
  {Class. Quant. Grav.}\ }\textbf {\bibinfo {volume} {30}},\ \bibinfo {pages}
  {135001} (\bibinfo {year} {2013})},\ \Eprint {http://arxiv.org/abs/1301.7419}
  {arXiv:1301.7419 [astro-ph.CO]} \BibitemShut {NoStop}%
\bibitem [{\citenamefont {Fadragas}\ \emph {et~al.}(2014)\citenamefont
  {Fadragas}, \citenamefont {Leon},\ and\ \citenamefont
  {Saridakis}}]{Fadragas:2013ina}%
  \BibitemOpen
  \bibfield  {author} {\bibinfo {author} {\bibfnamefont {Carlos~R.}\
  \bibnamefont {Fadragas}}, \bibinfo {author} {\bibfnamefont {Genly}\
  \bibnamefont {Leon}}, \ and\ \bibinfo {author} {\bibfnamefont {Emmanuel~N.}\
  \bibnamefont {Saridakis}},\ }\bibfield  {title} {\enquote {\bibinfo {title}
  {{Dynamical analysis of anisotropic scalar-field cosmologies for a wide range
  of potentials}},}\ }\href {\doibase 10.1088/0264-9381/31/7/075018} {\bibfield
   {journal} {\bibinfo  {journal} {Class. Quant. Grav.}\ }\textbf {\bibinfo
  {volume} {31}},\ \bibinfo {pages} {075018} (\bibinfo {year} {2014})},\
  \Eprint {http://arxiv.org/abs/1308.1658} {arXiv:1308.1658 [gr-qc]}
  \BibitemShut {NoStop}%
\bibitem [{\citenamefont {Skugoreva}\ \emph {et~al.}(2015)\citenamefont
  {Skugoreva}, \citenamefont {Saridakis},\ and\ \citenamefont
  {Toporensky}}]{Skugoreva:2014ena}%
  \BibitemOpen
  \bibfield  {author} {\bibinfo {author} {\bibfnamefont {Maria~A.}\
  \bibnamefont {Skugoreva}}, \bibinfo {author} {\bibfnamefont {Emmanuel~N.}\
  \bibnamefont {Saridakis}}, \ and\ \bibinfo {author} {\bibfnamefont
  {Alexey~V.}\ \bibnamefont {Toporensky}},\ }\bibfield  {title} {\enquote
  {\bibinfo {title} {{Dynamical features of scalar-torsion theories}},}\ }\href
  {\doibase 10.1103/PhysRevD.91.044023} {\bibfield  {journal} {\bibinfo
  {journal} {Phys. Rev. D}\ }\textbf {\bibinfo {volume} {91}},\ \bibinfo
  {pages} {044023} (\bibinfo {year} {2015})},\ \Eprint
  {http://arxiv.org/abs/1412.1502} {arXiv:1412.1502 [gr-qc]} \BibitemShut
  {NoStop}%
\bibitem [{\citenamefont {Dutta}\ \emph {et~al.}(2016)\citenamefont {Dutta},
  \citenamefont {Khyllep},\ and\ \citenamefont {Tamanini}}]{Dutta:2016bbs}%
  \BibitemOpen
  \bibfield  {author} {\bibinfo {author} {\bibfnamefont {Jibitesh}\
  \bibnamefont {Dutta}}, \bibinfo {author} {\bibfnamefont {Wompherdeiki}\
  \bibnamefont {Khyllep}}, \ and\ \bibinfo {author} {\bibfnamefont {Nicola}\
  \bibnamefont {Tamanini}},\ }\bibfield  {title} {\enquote {\bibinfo {title}
  {{Cosmological dynamics of scalar fields with kinetic corrections: Beyond the
  exponential potential}},}\ }\href {\doibase 10.1103/PhysRevD.93.063004}
  {\bibfield  {journal} {\bibinfo  {journal} {Phys. Rev. D}\ }\textbf {\bibinfo
  {volume} {93}},\ \bibinfo {pages} {063004} (\bibinfo {year} {2016})},\
  \Eprint {http://arxiv.org/abs/1602.06113} {arXiv:1602.06113 [gr-qc]}
  \BibitemShut {NoStop}%
\bibitem [{\citenamefont {Dutta}\ \emph {et~al.}(2017)\citenamefont {Dutta},
  \citenamefont {Khyllep},\ and\ \citenamefont {Tamanini}}]{Dutta:2017kch}%
  \BibitemOpen
  \bibfield  {author} {\bibinfo {author} {\bibfnamefont {Jibitesh}\
  \bibnamefont {Dutta}}, \bibinfo {author} {\bibfnamefont {Wompherdeiki}\
  \bibnamefont {Khyllep}}, \ and\ \bibinfo {author} {\bibfnamefont {Nicola}\
  \bibnamefont {Tamanini}},\ }\bibfield  {title} {\enquote {\bibinfo {title}
  {{Scalar-Fluid interacting dark energy: cosmological dynamics beyond the
  exponential potential}},}\ }\href {\doibase 10.1103/PhysRevD.95.023515}
  {\bibfield  {journal} {\bibinfo  {journal} {Phys. Rev. D}\ }\textbf {\bibinfo
  {volume} {95}},\ \bibinfo {pages} {023515} (\bibinfo {year} {2017})},\
  \Eprint {http://arxiv.org/abs/1701.00744} {arXiv:1701.00744 [gr-qc]}
  \BibitemShut {NoStop}%
\bibitem [{\citenamefont {Zonunmawia}\ \emph {et~al.}(2018)\citenamefont
  {Zonunmawia}, \citenamefont {Khyllep}, \citenamefont {Dutta},\ and\
  \citenamefont {J\"arv}}]{Zonunmawia:2018xvf}%
  \BibitemOpen
  \bibfield  {author} {\bibinfo {author} {\bibfnamefont {Hmar}\ \bibnamefont
  {Zonunmawia}}, \bibinfo {author} {\bibfnamefont {Wompherdeiki}\ \bibnamefont
  {Khyllep}}, \bibinfo {author} {\bibfnamefont {Jibitesh}\ \bibnamefont
  {Dutta}}, \ and\ \bibinfo {author} {\bibfnamefont {Laur}\ \bibnamefont
  {J\"arv}},\ }\bibfield  {title} {\enquote {\bibinfo {title} {{Cosmological
  dynamics of brane gravity: A global dynamical system perspective}},}\ }\href
  {\doibase 10.1103/PhysRevD.98.083532} {\bibfield  {journal} {\bibinfo
  {journal} {Phys. Rev. D}\ }\textbf {\bibinfo {volume} {98}},\ \bibinfo
  {pages} {083532} (\bibinfo {year} {2018})},\ \Eprint
  {http://arxiv.org/abs/1810.03816} {arXiv:1810.03816 [gr-qc]} \BibitemShut
  {NoStop}%
\bibitem [{\citenamefont {Khyllep}\ and\ \citenamefont
  {Dutta}(2021)}]{Khyllep:2021yyp}%
  \BibitemOpen
  \bibfield  {author} {\bibinfo {author} {\bibfnamefont {Wompherdeiki}\
  \bibnamefont {Khyllep}}\ and\ \bibinfo {author} {\bibfnamefont {Jibitesh}\
  \bibnamefont {Dutta}},\ }\bibfield  {title} {\enquote {\bibinfo {title}
  {{Cosmological dynamics and bifurcation analysis of the general non-minimal
  coupled scalar field models}},}\ }\href {\doibase
  10.1140/epjc/s10052-021-09559-x} {\bibfield  {journal} {\bibinfo  {journal}
  {Eur. Phys. J. C}\ }\textbf {\bibinfo {volume} {81}},\ \bibinfo {pages} {774}
  (\bibinfo {year} {2021})},\ \Eprint {http://arxiv.org/abs/2102.04744}
  {arXiv:2102.04744 [gr-qc]} \BibitemShut {NoStop}%
\bibitem [{\citenamefont {Basilakos}\ \emph {et~al.}(2019)\citenamefont
  {Basilakos}, \citenamefont {Leon}, \citenamefont {Papagiannopoulos},\ and\
  \citenamefont {Saridakis}}]{Basilakos:2019dof}%
  \BibitemOpen
  \bibfield  {author} {\bibinfo {author} {\bibfnamefont {Spyros}\ \bibnamefont
  {Basilakos}}, \bibinfo {author} {\bibfnamefont {Genly}\ \bibnamefont {Leon}},
  \bibinfo {author} {\bibfnamefont {G.}~\bibnamefont {Papagiannopoulos}}, \
  and\ \bibinfo {author} {\bibfnamefont {Emmanuel~N.}\ \bibnamefont
  {Saridakis}},\ }\bibfield  {title} {\enquote {\bibinfo {title} {{Dynamical
  system analysis at background and perturbation levels: Quintessence in severe
  disadvantage comparing to $\Lambda$CDM}},}\ }\href {\doibase
  10.1103/PhysRevD.100.043524} {\bibfield  {journal} {\bibinfo  {journal}
  {Phys. Rev. D}\ }\textbf {\bibinfo {volume} {100}},\ \bibinfo {pages}
  {043524} (\bibinfo {year} {2019})},\ \Eprint
  {http://arxiv.org/abs/1904.01563} {arXiv:1904.01563 [gr-qc]} \BibitemShut
  {NoStop}%
\bibitem [{\citenamefont {Alho}\ \emph {et~al.}(2019)\citenamefont {Alho},
  \citenamefont {Uggla},\ and\ \citenamefont {Wainwright}}]{Alho:2019jho}%
  \BibitemOpen
  \bibfield  {author} {\bibinfo {author} {\bibfnamefont {Artur}\ \bibnamefont
  {Alho}}, \bibinfo {author} {\bibfnamefont {Claes}\ \bibnamefont {Uggla}}, \
  and\ \bibinfo {author} {\bibfnamefont {John}\ \bibnamefont {Wainwright}},\
  }\bibfield  {title} {\enquote {\bibinfo {title} {{Perturbations of the
  Lambda-CDM model in a dynamical systems perspective}},}\ }\href {\doibase
  10.1088/1475-7516/2019/09/045} {\bibfield  {journal} {\bibinfo  {journal}
  {JCAP}\ }\textbf {\bibinfo {volume} {09}},\ \bibinfo {pages} {045} (\bibinfo
  {year} {2019})},\ \Eprint {http://arxiv.org/abs/1904.02463} {arXiv:1904.02463
  [gr-qc]} \BibitemShut {NoStop}%
\bibitem [{\citenamefont {Landim}(2019)}]{Landim:2019lvl}%
  \BibitemOpen
  \bibfield  {author} {\bibinfo {author} {\bibfnamefont {Ricardo~G.}\
  \bibnamefont {Landim}},\ }\bibfield  {title} {\enquote {\bibinfo {title}
  {{Cosmological perturbations and dynamical analysis for interacting
  quintessence}},}\ }\href {\doibase 10.1140/epjc/s10052-019-7418-8} {\bibfield
   {journal} {\bibinfo  {journal} {Eur. Phys. J. C}\ }\textbf {\bibinfo
  {volume} {79}},\ \bibinfo {pages} {889} (\bibinfo {year} {2019})},\ \Eprint
  {http://arxiv.org/abs/1908.03657} {arXiv:1908.03657 [gr-qc]} \BibitemShut
  {NoStop}%
\bibitem [{\citenamefont {Khyllep}\ \emph {et~al.}(2022)\citenamefont
  {Khyllep}, \citenamefont {Dutta}, \citenamefont {Basilakos},\ and\
  \citenamefont {Saridakis}}]{Khyllep:2021wjd}%
  \BibitemOpen
  \bibfield  {author} {\bibinfo {author} {\bibfnamefont {Wompherdeiki}\
  \bibnamefont {Khyllep}}, \bibinfo {author} {\bibfnamefont {Jibitesh}\
  \bibnamefont {Dutta}}, \bibinfo {author} {\bibfnamefont {Spyros}\
  \bibnamefont {Basilakos}}, \ and\ \bibinfo {author} {\bibfnamefont
  {Emmanuel~N.}\ \bibnamefont {Saridakis}},\ }\bibfield  {title} {\enquote
  {\bibinfo {title} {{Background evolution and growth of structures in
  interacting dark energy scenarios through dynamical system analysis}},}\
  }\href {\doibase 10.1103/PhysRevD.105.043511} {\bibfield  {journal} {\bibinfo
   {journal} {Phys. Rev. D}\ }\textbf {\bibinfo {volume} {105}},\ \bibinfo
  {pages} {043511} (\bibinfo {year} {2022})},\ \Eprint
  {http://arxiv.org/abs/2111.01268} {arXiv:2111.01268 [gr-qc]} \BibitemShut
  {NoStop}%
\bibitem [{\citenamefont {Song}\ \emph {et~al.}(2007)\citenamefont {Song},
  \citenamefont {Hu},\ and\ \citenamefont {Sawicki}}]{Song:2006ej}%
  \BibitemOpen
  \bibfield  {author} {\bibinfo {author} {\bibfnamefont {Yong-Seon}\
  \bibnamefont {Song}}, \bibinfo {author} {\bibfnamefont {Wayne}\ \bibnamefont
  {Hu}}, \ and\ \bibinfo {author} {\bibfnamefont {Ignacy}\ \bibnamefont
  {Sawicki}},\ }\bibfield  {title} {\enquote {\bibinfo {title} {{The Large
  Scale Structure of f(R) Gravity}},}\ }\href {\doibase
  10.1103/PhysRevD.75.044004} {\bibfield  {journal} {\bibinfo  {journal} {Phys.
  Rev. D}\ }\textbf {\bibinfo {volume} {75}},\ \bibinfo {pages} {044004}
  (\bibinfo {year} {2007})},\ \Eprint {http://arxiv.org/abs/astro-ph/0610532}
  {arXiv:astro-ph/0610532} \BibitemShut {NoStop}%
\bibitem [{\citenamefont {Sagredo}\ \emph {et~al.}(2018)\citenamefont
  {Sagredo}, \citenamefont {Nesseris},\ and\ \citenamefont
  {Sapone}}]{Sagredo:2018ahx}%
  \BibitemOpen
  \bibfield  {author} {\bibinfo {author} {\bibfnamefont {Bryan}\ \bibnamefont
  {Sagredo}}, \bibinfo {author} {\bibfnamefont {Savvas}\ \bibnamefont
  {Nesseris}}, \ and\ \bibinfo {author} {\bibfnamefont {Domenico}\ \bibnamefont
  {Sapone}},\ }\bibfield  {title} {\enquote {\bibinfo {title} {{Internal
  Robustness of Growth Rate data}},}\ }\href {\doibase
  10.1103/PhysRevD.98.083543} {\bibfield  {journal} {\bibinfo  {journal} {Phys.
  Rev. D}\ }\textbf {\bibinfo {volume} {98}},\ \bibinfo {pages} {083543}
  (\bibinfo {year} {2018})},\ \Eprint {http://arxiv.org/abs/1806.10822}
  {arXiv:1806.10822 [astro-ph.CO]} \BibitemShut {NoStop}%
\bibitem [{\citenamefont {Kazantzidis}\ and\ \citenamefont
  {Perivolaropoulos}(2018)}]{Kazantzidis:2018rnb}%
  \BibitemOpen
  \bibfield  {author} {\bibinfo {author} {\bibfnamefont {Lavrentios}\
  \bibnamefont {Kazantzidis}}\ and\ \bibinfo {author} {\bibfnamefont
  {Leandros}\ \bibnamefont {Perivolaropoulos}},\ }\bibfield  {title} {\enquote
  {\bibinfo {title} {{Evolution of the $f\sigma_8$ tension with the
  Planck15/$\Lambda$CDM determination and implications for modified gravity
  theories}},}\ }\href {\doibase 10.1103/PhysRevD.97.103503} {\bibfield
  {journal} {\bibinfo  {journal} {Phys. Rev. D}\ }\textbf {\bibinfo {volume}
  {97}},\ \bibinfo {pages} {103503} (\bibinfo {year} {2018})},\ \Eprint
  {http://arxiv.org/abs/1803.01337} {arXiv:1803.01337 [astro-ph.CO]}
  \BibitemShut {NoStop}%
\end{thebibliography}%
	
\end{document}